%% file: main.tex
\documentclass[11pt]{article}

\usepackage[margin=1in]{geometry}
\usepackage{amsmath,amssymb}
\usepackage{graphicx}
\usepackage{booktabs}
\usepackage[numbers,sort&compress]{natbib}
\usepackage{url}
\usepackage{tikz}
\usetikzlibrary{arrows.meta,positioning,fit,calc,backgrounds}
\usepackage[hidelinks]{hyperref}

\graphicspath{{./}}

\ExplSyntaxOn
\NewDocumentCommand{\code}{m}
  {
    \texttt
      {
        \tl_set:Nn \l_tmpa_tl {#1}
        \tl_replace_all:Nnn \l_tmpa_tl { \_ } { \_\allowbreak }
        \tl_replace_all:Nnn \l_tmpa_tl { / } { /\allowbreak }
        \tl_replace_all:Nnn \l_tmpa_tl { = } { =\allowbreak }
        \tl_replace_all:Nnn \l_tmpa_tl { . } { .\allowbreak }
        \tl_use:N \l_tmpa_tl
      }
  }
\ExplSyntaxOff

\title{Safety-Gated Agentic Supervisory Control on a Coupled Distillation
Benchmark:\\[0.25em]
\large Regime Map, Auditable Gate, and Co-Design Findings}
\author{Christian Rosenthal\\[0.35em]
\small Sole author --- independent research, industrial AI methodology\\
\url{https://github.com/cgncro-cyber/IndustrialAI}}
\date{Preprint --- July 2026}

\begin{document}
\maketitle

\begin{abstract}
An open-weight LLM can write composition setpoints every five minutes. What a
plant still needs is a hard check: named constraints, logged margins, and an
admit/block decision before the regulatory layer moves. This paper puts that
check in a rule-based forked-twin counterfactual gate (nine pinned
constraints) and leaves the regulatory layer unchanged. On Skogestad's
Column~A the ladder is PID-only~(C0), linear MPC~(C1), ungated agent~(C2),
and gated agent~(C3) under one contract---identical level closure ($M_D$,
$M_B$), scenarios, and seeds; C2/C3 share the linear-MPC backend.

The split is not subtle. Off-nominal \emph{target acquisition}: the agent
beats Pareto-tuned linear MPC in the strong band (C2/C1 IAE ratio $0.361$ at
the upper CI). \emph{Disturbance rejection} on the same 16-point grid: inverse
by $16.03$ at the upper CI ($10.18$ at the point estimate). That is where an
ungated LLM supervisor does not belong. The gate compresses a
specification-abandonment attractor into a bounded offset ($d\approx -1.4$;
P95 cell IAE $11.5\rightarrow 0.77$). A one-line prompt fix kills the
attractor at the source ($6/10\rightarrow 0/10$; sensitivity only---not a new
headline). In a 250-cell statistical pass, $534$ of $590$ gate interventions
are spec-on-bound geometry: the operating specification sits on a safety
limit, so a well-behaved OP becomes inoperable while misbehaving ones are
only contained. The same gate still corrects $318$ actively harmful
proposals.

Headlines are single-column and model-conditional on DeepSeek-V4-Flash. A full
second-family sweep (NVIDIA Nemotron-3-Super) keeps the disturbance-rejection
\code{fails} band and the plant-side failure geography; ratio magnitude and
protocol operability stay model-conditional, and Super target-acquisition
``strong'' cells are survivors only (not confirmation). Transfer means twin,
constraint envelope, and setpoint interface---not a second plant class
measured here.
\end{abstract}

\section{Introduction}\label{sec:intro}

Supervisory control is a practical place to put a large language model.
Regimes shift, retuning a model-based controller is expensive, and a
five-to-fifteen minute cadence leaves room for inference. It is also a place
where an unaudited black box will not pass a HAZOP conversation. A setpoint
pair can look reasonable and still be plant-unsafe a few cycles later.

So the useful question is not ``can an LLM emit targets?'' It is whether
every proposal is checked, every intervention names a constraint and a
margin, and failure modes are measured instead of assumed away.

What follows is organized around four results.

\begin{enumerate}
\item \textbf{Architecture.}
 A deterministic Observer--Optimizer--Critic loop in plain Python (no
 third-party graph runtime) over an unchanged regulatory layer. Between
 proposal and execution sits a forked-twin counterfactual gate: 30-minute
 horizon, 5-minute fast-fail, nine pinned constraints. Each gate decision
 is reproducible from the twin, the constraint list, and a documented
 numerical tolerance.
\item \textbf{Four-way ladder.}
 C0 PID-only, C1 linear MPC, C2 agent, C3 gated agent---same level closure,
 scenarios, and seeds; C2/C3 on the same MPC backend (design records in
 Appendix~\ref{app:records}). That split isolates the agent's marginal
 value from the gate's.
\item \textbf{Gate co-design.}
 Containment without recovery on a specification-abandonment attractor;
 the onset-invisible / tail-intercepted boundary of single-proposal
 trajectory checks; and a geometry-dominated intercept volume---$534$ of
 $590$ interventions in a 250-cell pass sit on a safety bound. Specification,
 envelope, and admission rule have to be designed together.
\item \textbf{Regime map.}
 Strong value on off-nominal \emph{target acquisition} under linearization
 stress; negative value on steady-state \emph{disturbance rejection}
 (\S\ref{sec:bucket}, \S\ref{sec:failuregeo}). Practical reading: MPC
 default for local regulation; gated LLM only where re-planning is
 warranted---not wholesale MPC replacement (\S\ref{sec:hybridfw}).
\end{enumerate}

\paragraph{Scope.}
Headlines use one canonical column (Skogestad Column~A) and are
model-conditional on DeepSeek-V4-Flash (classification frozen 2026-07-02;
Appendix~\ref{app:records}). The completed Nemotron-3-Super sweep keeps the
disturbance-rejection \code{fails} band and the plant-side failure geography,
with model-conditional magnitude and survivorship-qualified
target-acquisition survivors (\S\ref{sec:crossfamily}). Transfer is
architectural, not empirical (\S\ref{sec:discussion}). A nominal-only
MPC-free branch shows agent-over-PID is operable ($50/50$ cells) but sits
below PID-only on every disturbance arm (\S\ref{sec:mpcfree}).

\section{Related Work}\label{sec:related}

\paragraph{Classical multivariable and supervisory process control.}
Distillation dual-composition control is a canonical multivariable problem
(RGA analysis, decentralized PID vs model-predictive control, two-layer
supervisory/regulatory hierarchies) \cite{qin2003survey, skogestad2004control,
tatjewski2008advanced}. Our C0 and C1 baselines instantiate this lineage
deliberately --- relay-tuned decentralized PID and a linear MPC with a
regularization Pareto sweep --- so that the agent is compared against the
industrial standard, not a strawman.

\paragraph{Reinforcement learning for process control.}
Learned-policy approaches \cite{nian2020review, spielberg2019toward} adapt
across regimes but concentrate their evidence in the policy's weights;
auditability is post-hoc at best. We position the LLM-agent paradigm
differently: the supervisor emits a structured, rationale-carrying proposal
per cycle, and the safety argument is carried by a deterministic gate outside
the learned component rather than by properties of the policy itself.

\paragraph{Agentic LLM systems.}
The Observer/Optimizer/Critic decomposition follows the tool-calling agent
lineage (structured observation, single LLM decision step, rule-based
validation with bounded revision) --- a lineage spanning autonomous chemistry
agents \cite{boiko2023autonomous, bran2024augmenting}, LLM-orchestrated
production systems \cite{xia2023towards, gill2025leveraging}, and
LLM-driven control design \cite{guo2024controlagent}. Our contribution to this line is not the loop
but its evaluation discipline: pinned sampling, seeds, prompts, and
fail-fast semantics, evaluated at statistical scale against model-based
baselines.

\paragraph{Anomaly detection on industrial benchmarks.}
Benchmark-trained detectors (e.g., on the Tennessee Eastman process
\cite{downs1993plantwide, venkatasubramanian2003review, yin2012comparison})
are the natural learned complement to our gate. In this study the learned
layer was deliberately descoped (the safety-layer descope, Appendix~\ref{app:records}): its cross-plant feature mapping
was never satisfactorily specified, and the rule-based counterfactual gate
alone carries the safety contribution. Learned early-warning layers on top
of the guaranteed gate remain future work.

\paragraph{Positioning against the closest prior.}
Nogueira \& Skogestad \cite{nogueira2026systematic} place LLM operator
agents \emph{inside} the regulatory layer: each loop of an
advanced-regulatory-control chain is mapped to one specialized agent, and
safety is obtained \emph{structurally} --- a deterministic selector-priority
orchestrator resolves every constraint conflict regardless of the LLM output
--- demonstrated on a single ventilation case study. Our architecture is
complementary on all three axes: the LLM acts at the \emph{supervisory}
layer (composition setpoints over an unchanged regulatory layer), safety is
enforced \emph{behaviorally} by a counterfactual forked-twin simulation gate
rather than structurally by priority wiring, and the evaluation is
statistical --- $N = 10$ seeds per cell, a 250-cell pass --- against
PID-only, linear-MPC, and ungated-agent baselines. Vyas \& Mercang\"oz
\cite{vyas2025autonomous} likewise couple LLM agents with a digital twin for
risk-free pre-execution assessment and name twin-grounded reasoning about
``constraint feasibility, and what-if counterfactuals'' as the enabling
direction; our gate operationalizes exactly that direction as a pinned
nine-constraint admission rule.

\paragraph{Gap.}
The gate sits in the Simplex / shielding / runtime-assurance line
\cite{sha2001simplicity, seto1998simplex, alshiekh2018safe,
hobbs2023runtime}: a trusted simple mechanism around an untrusted complex
controller, here at supervisory cadence with trajectory (not action-set)
semantics. Relative to regulatory-layer multi-agent orchestration with
structural priority safety \cite{nogueira2026systematic} and twin-grounded
LLM risk assessment \cite{vyas2025autonomous}, three pieces are joined that
are usually separated: a supervisory-layer LLM, a behavioral forked-twin
gate with named margins, and a pre-registered four-way statistical ladder
that treats the inverse disturbance-rejection result and the gate's own
failure surface (onset invisibility, spec-on-bound inadmissibility) as
results, not caveats. That combination on a canonical multivariable
benchmark is the engineering gap this study fills.

\section{Methodology}\label{sec:methodology}

\subsection{Process twin}\label{sec:twin}

All experiments run on a dynamic Python port of Skogestad's ``Column~A''
\cite{skogestad1988understanding, skogestad1996multivariable,
skogestad1997dynamics}: a 40-stage binary distillation column in the LV
configuration at the nominal operating point $F = 1$~kmol/min, $z_F = 0.5$,
$q_F = 1$, integrated with \code{scipy.integrate.\allowbreak solve\_ivp}
\cite{virtanen2020scipy} over an 82-dimensional state (41 stage compositions,
41 holdups) and validated against published NTNU reference trajectories.
Column~A is the canonical multivariable distillation benchmark and is
genuinely coupled: RGA$(1,1)\approx 36$ at the nominal operating point
\cite{bristol1966new}, so dual-composition control is not two independent
loops (\S\ref{sec:limitations}). DV or L/D-V/B would decouple better; LV is
retained as the literature comparison configuration and because the
regulatory layer must stay fixed across C0--C3. Condenser and reboiler
holdups ($M_D$, $M_B$) under proportional control are \textbf{identical
across all four configurations}---the foundation of the apples-to-apples
contract: same plant, same inventory dynamics, same disturbance path.

\subsection{Four-way comparison contract}\label{sec:contract}

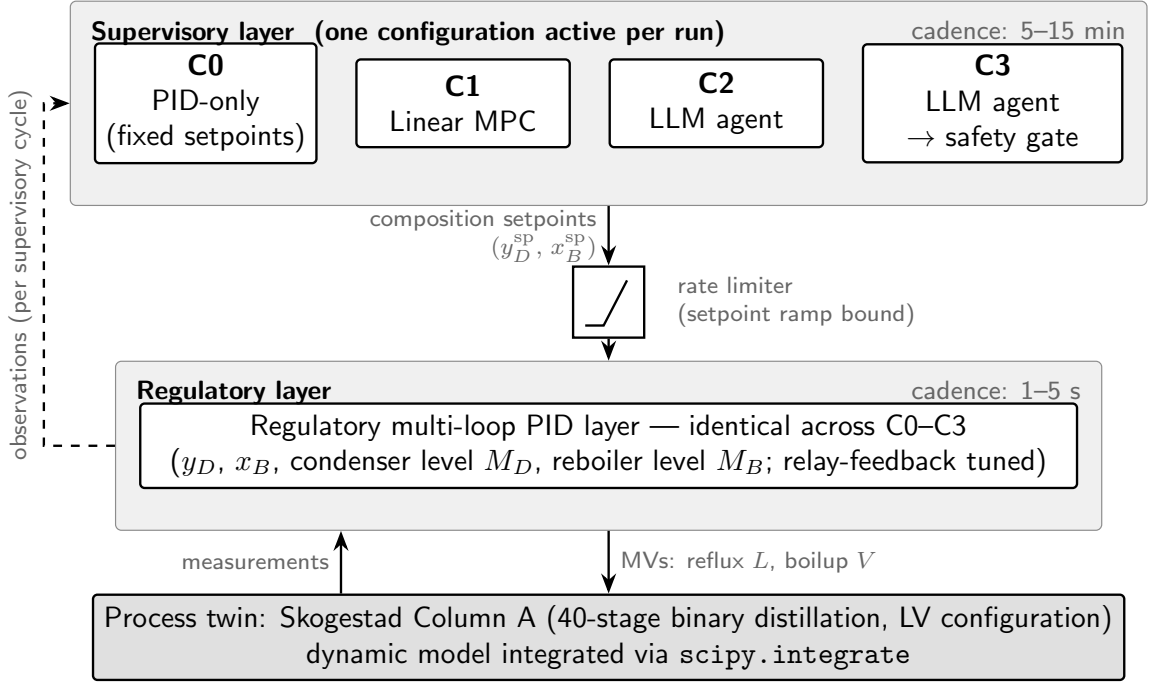
\begin{figure}[!htbp]
 \centering
 \vspace{0pt}%
 \resizebox{0.95\textwidth}{!}{%
   \input{fig1_architecture_body}%
 }%
 \caption{Two-layer hierarchy. The variable under study is the supervisory
 policy (C0--C3). Composition setpoints pass a rate limiter into a
 regulatory layer that is held fixed on the primary ladder and actuates the
 Column~A twin. \emph{Takeaway:} the LLM never touches actuators.}
 \label{fig:arch}
\end{figure}

Figure~\ref{fig:arch} holds everything constant except the supervisory
mechanism: \textbf{C0}---decentralized PID composition control with fixed
operator setpoints; \textbf{C1}---linear MPC replacing the composition PIDs;
\textbf{C2}---the agentic supervisor proposing composition setpoints;
\textbf{C3}---C2 plus the counterfactual gate. Level closure, scenarios, and
seeds are identical. C2 and C3 use the \textbf{same MPC regulatory backend},
so C3$-$C2 isolates the gate and C2$-$C1 isolates the supervisor. A
PID-backend ``MPC-free'' branch is secondary analysis at demonstration
scope---nominal regime only, descriptive, no outcome classification
(\S\ref{sec:mpcfree}; design records in Appendix~\ref{app:records}).

\subsection{Baseline controllers C0 and C1}\label{sec:baselines}

\paragraph{C0 provenance.}
The PID baseline is not a strawman: its tuning was selected by a
pre-registered shootout (relay-feedback identification
\cite{astrom1984automatic}, Tyreus--Luyben synthesis \cite{tyreus1992tuning},
decoupler and detuning variants), won by the \code{TL\_no\_decoupler}
variant with aggregate IAE 0.836 mole-fraction$\cdot$min over the canonical
scenario set --- a 29\,\% margin over the runner-up
(versioned shootout artifact in the public repository). Its fixed gains deliberately do not
extrapolate to $F \pm 20\,\%$ operating points (\S\ref{sec:limitations})
--- the expected property of single-OP tuning, and the first half of the
motivation for supervisory adaptation.

\begin{figure}[tbp]
 \centering
 \includegraphics[width=\textwidth]{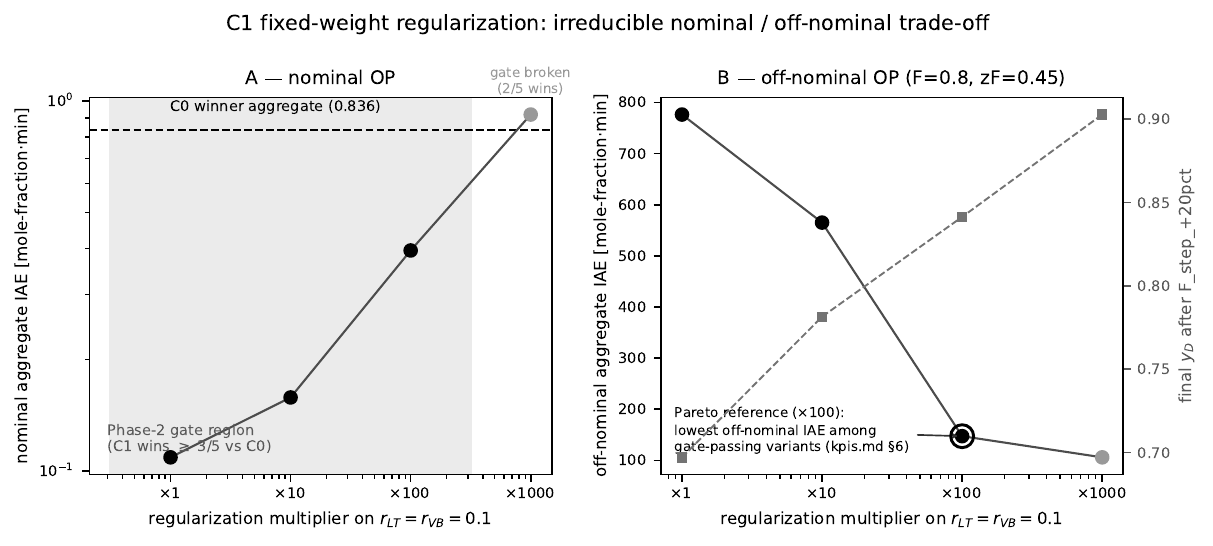}
 \caption{C1 fixed-weight regularization Pareto front. Panel A: nominal
 aggregate IAE vs multiplier with the C0 winner aggregate reference and
 the shaded baseline gate region (C1 wins $\geq 3/5$ vs C0); the
 gate-breaking $\times 1000$ variant is greyed out. Panel B: off-nominal
 aggregate IAE at the worst-cluster OP $(F{=}0.8, z_F{=}0.45)$ with the
 final $y_D$ after \code{F\_step\_+20pct} on the secondary axis; the
 $\times 100$ gate-passing variant is the outcome-class-B Pareto reference
. All values read from the versioned sweep artifact;
 nothing recomputed.}
 \label{fig:pareto}
\end{figure}

\paragraph{C1 and the ``why an agent at all'' argument.}
The MPC baseline linearizes the LV-closed plant per operating point from a
cached operating-window sweep and solves a QP at the supervisory cadence
(implemented with do-mpc \cite{fiedler2023dompc} on CasADi
\cite{andersson2019casadi}); at the nominal OP it beats C0 on 5/5 scenarios
(aggregate 0.122 vs 0.836; the linear-MPC-baseline gate). Its structural limit is the
regularization trade-off, reported as a complete Pareto front rather than a
single tuning (Figure~\ref{fig:pareto}): across multipliers
$\times 1$--$\times 1000$ on the MV-move penalties, \textbf{no fixed-weight
variant wins both regimes} --- the nominal-best tuning collapses
off-nominal, and the strongest gate-passing off-nominal variant
($\times 100$, the Pareto reference all C2 comparisons use) still fails the
low-$F$ corner where the LV gain matrix approaches rank deficiency
($\mathrm{cond}(G_{mv})$ $150 \rightarrow 6800$ across the grid;
\S\ref{sec:limitations}). This irreducible trade-off is the structural
opening for an adaptive supervisory layer: a supervisor that re-plans
targets when the linearization is stressed addresses exactly the regime no
fixed-weight linear MPC covers. The agent is evaluated against the Pareto
reference, not against the nominal-best variant, to pre-empt the
strawman-MPC objection.

\subsection{Agentic supervisor (C2)}\label{sec:agent}

\subsubsection{Architecture}\label{sec:agentarch}

{\sloppy
Each five-minute cycle is an Observer $\rightarrow$ Optimizer $\rightarrow$
Critic loop---sense, decide, validate, actuate---implemented in plain
Python so revision budgets and escalate paths stay inspectable. The
\emph{Observer} is deterministic: it packs compositions, MVs, running IAE,
and feed conditions into a structured prompt. The \emph{Optimizer} is the
only LLM call; it returns JSON keys \code{y\_D\_target},
\code{x\_B\_target}, and \code{rationale}, parsed strictly. The
\emph{Critic} is rule-based (not an LLM): physical bounds, ordering
(\code{y\_D\_target} must exceed \code{x\_B\_target}), and related checks.
A reject returns to the Optimizer with chain-of-thought enabled.\par}

The revision budget is hard-capped at two per cycle. Exhaustion with a prior
accepted target logs a documented \code{escalate} (re-emit last accepted).
Exhaustion on the first cycle of a run raises
\code{CriticLoop\-LimitError} and aborts---no silent default. Accepted
targets are slew-limited into the regulatory layer (\S\ref{sec:contract}).

\subsubsection{LLM specification}\label{sec:llmspec}

The \textbf{headline} model (frozen classification and gate statistical
pass) is DeepSeek-V4-Flash \cite{deepseekv4flash2026}
(\code{deepseek/\allowbreak deepseek-v4-flash}), served via OpenRouter to a
pinned provider (Novita, \code{allow\_fallbacks=false}). Distinguish
\emph{request mode} (API flags: first-round thinking/reasoning off under the
project modal policy) from \emph{realized mode} (what the host returns). On
this path the request is off, but Novita's chat template does not honour
thinking-off: responses still carry a chain-of-thought trace, so the
headline stack ran under \textbf{provider-realized always-on reasoning}
(provider-boundary finding; see the project reproducibility record
in Appendix~\ref{app:records}).
That is a provider-boundary fact about the V4-Flash evaluation host, not a
change to the Nemotron DoE pin. Cross-family contrasts therefore confound
model family, sampling temperature, \emph{and} reasoning realization
(\S\ref{sec:crossfamily}). The cross-family re-instantiation is
Nemotron-3-Super-120B-A12B \cite{nemotron3super2026} (120\,B total / 12\,B
active) at its DoE pin on a throttle-free paid endpoint (DeepInfra bf16;
first-round realized reasoning-off). Early screening used NVIDIA NIM; the
full 1600-cell Super sweep uses the DeepInfra pin. The model-selection
journey (including 49\,B multi-step-coherence failure) is in
Appendix~\ref{app:modelselection}.

\subsubsection{Inference configuration (DoE-derived sampling)}\label{sec:doe}

Sampling hyperparameters were selected, not assumed. A full-factorial design
of experiments on the nominal-baseline scenario --- temperature
$\in \{0.0, 0.3, 0.6, 0.8, 1.0\}$ $\times$ top\_p $\in \{0.8, 0.95, 1.0\}$
$\times$ reasoning configuration $\in$ \{off, on@1024, on@4096\}, 45 cells
$\times$ $N=5$ seeds $= 225$ screening runs --- identified a plateau of
near-zero canonical IAE for $T \leq 0.6$ with reasoning off, a performance
cliff at $T=0.8$, and an instability at $T=1.0$ that chain-of-thought
reasoning rescues. The Nemotron DoE pins $T=0.3$, top\_p $=0.95$,
reasoning off with \code{max\_tokens=512} on first-round calls and reasoning
on with \code{max\_tokens=4096} on Critic-revision calls (the modal
reasoning policy for that family). $T=0.3$ was chosen over $T=0.0$ (degenerative-repetition
risk on long sequences) and $T=0.6$ (thin margin to the $T=0.8$ cliff) as
the plateau-interior point with the largest empirical margin. A confirmation
run at the selected cell ($N=10$ seeds) yields mean canonical IAE
$5.75 \times 10^{-7}$ mole-fraction$\cdot$min --- roughly four orders of
magnitude below the 0.01 acceptance threshold, with all ten seed IAEs
identical to seven significant figures. The V4-Flash \emph{headline} sampling uses that family's own pin ($T=1.0$,
top\_p $=0.95$) with \emph{realized} always-on reasoning under the Novita
template (\S\ref{sec:llmspec})---distinct from Super's first-round
reasoning-off pin at $T=0.3$. Cross-family comparisons therefore contrast
each model at its validated operating point rather than a shared temperature
or shared realized reasoning mode (\S\ref{sec:crossfamily}). Sampling
parameters are pinned in source, not user-configurable; response-surface
detail is in Appendix~\ref{app:doe}.

\subsubsection{Reproducibility chain}\label{sec:repro}

The agentic layer is reproducibly specified across six axes: (i) model
weights, open on HuggingFace; (ii) chat template, public in the model
repository; (iii) sampling parameters, pinned in source (\S\ref{sec:doe});
(iv) seeds, pinned per scenario in the public drivers; (v) inference
protocol, the open OpenAI Chat Completions standard; (vi) inference host: a
hosted, per-token, no-NDA OpenAI-compatible endpoint (not a free-tier
availability claim---free-tier throttling previously aborted a NIM sweep;
the V4-Flash headline used paid/pinned OpenRouter routing). Two independent
reproduction paths follow: a hosted replication (same endpoint and model
identifiers) and a self-hosted replication (deploy published weights via
vLLM, SGLang, or TensorRT-LLM and swap the base URL). Closed-weight
commercial APIs degrade at least three of these axes simultaneously ---
deprecation cycles erase the named version, silent model updates perturb
behavior between calls, and opaque host templates raise fidelity risk. The architecture result is claimed at the level of \emph{an open-weight
agentic-post-trained LLM, reproducibly pinned}; the cross-family ablation on
the same chain supports that the claim is not specific to one model family.

\subsubsection{Failure-mode discipline}\label{sec:failfast}

The agentic layer fails fast rather than degrading silently.
Transport failures (timeouts, connection resets, non-2xx responses) raise
named exceptions and terminate the loop, with the cycle marked failed in the
audit log --- no retry loops, no fallback model, no fallback provider. Parse
failures trigger a Critic revision with diagnostic context; on budget
exhaustion the \S\ref{sec:agentarch} semantics apply (logged, counted
\code{escalate} when a prior accepted target exists;
\code{CriticLoopLimitError} when none does). Configuration failures (missing
credentials, unknown model identifier, unknown reasoning protocol) raise at
client-construction time, before any cycle runs. The mock LLM client used in
unit tests cannot be constructed outside a test context. These guarantees
mean every cycle in the evaluation dataset was produced by the exact
configuration named in its manifest --- the property the outcome
classification's integrity rests on.

\subsection{Safety gate (C3)}\label{sec:gate}

\subsubsection{Counterfactual gate (primary mechanism)}\label{sec:gatemech}

\begin{figure}[tbp]
 \centering
 \includegraphics[width=0.9\textwidth]{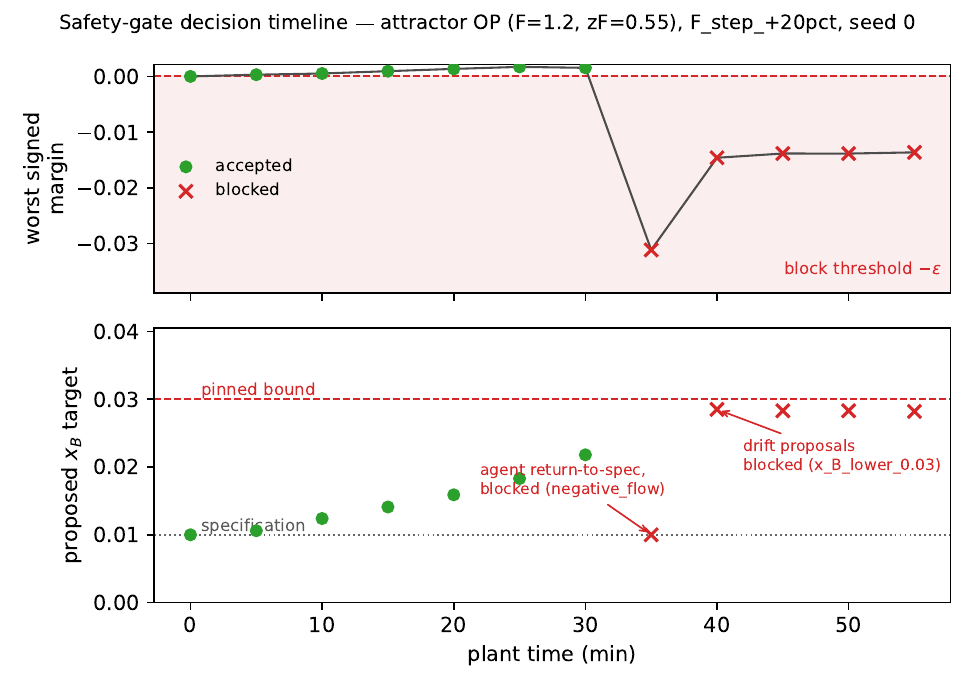}
 \caption{Safety-gate decision timeline for the reference cell (attractor
 OP, seed 0, 2026-07-06 evaluation): worst signed constraint margin of each
 gated proposal's counterfactual with the block threshold
 $-\varepsilon$ marked (top), and proposed $x_B$ targets with
 accepted/blocked verdicts, trigger constraints, and the blocked recovery
 proposal (cs02) annotated (bottom).}
 \label{fig:gatetimeline}
\end{figure}

The C3 configuration wraps the C2 supervisor's proposal interface in a
rule-based forked-twin counterfactual gate: at every supervisory proposal
the plant state is deep-copied, integrated 30 minutes forward (six
supervisory cycles) under the proposed setpoints with the regulatory layer
running normally and the feed disturbance held at its fork-time value, and
the resulting trajectory is tested against the nine pre-registered safety
constraints (distillate- and bottoms-purity bounds, condenser/reboiler
holdup bands, composition and flow physical-feasibility bounds, and the
actuator envelope). Violations inside the first five minutes are flagged
separately (fast-fail: immediate-harm proposals no supervisory cycle could
avert). On a violation the gate blocks the proposal, substitutes the last
gate-accepted target as the documented safe action, and writes a structured
log row (trigger constraint, signed margins for all nine constraints,
rejected proposal, substitute) that feeds the intercept-rate accounting
(Figure~\ref{fig:gatetimeline}); an inconclusive counterfactual (integration
failure) blocks conservatively under its own trigger class. The violation
decision carries a float-noise tolerance of $10^{-6}$ on the signed margin
--- adopted mid-evaluation with documented justification when on-spec
initial states (steady-state solver residuals of order $10^{-8}$ on a
specification that sits exactly on the purity bound) were classified as
violated at $t = 0$; the pinned threshold values are unchanged, logged
margins stay raw, and the tolerance is result-independent and
direction-blind by construction and by test. There is no learned component
in the loop: every gate decision is exactly reproducible from the twin, the
constraint list, and the tolerance.

\subsubsection{Intercept and detection KPIs}\label{sec:gatekpis}

On the first evaluation pass (four operating points $\times$ the
\code{F\_step\_+20pct} disturbance $\times$ ten seeds, DeepSeek-V4-Flash
under the fixed v1 prompt): 0/120 intercepts at the nominal reference (the
gate's false-positive check under the tolerance), 23/120 at the
$x_B$-channel follower OP and 25/120 at the $y_D$-channel follower OP ---
all fast-fail, i.e.\ the counterfactual verdict was decided within the first
five minutes of the fork --- and a deterministic first-cycle block at the
fourth OP (\S\ref{sec:scopeboundary} and \S\ref{sec:discussion}).
Worst-case cell IAE at the follower OPs drops from 11.46 (C2) to 0.774 and
from 9.48 to 0.604; cell medians from 6.40 and 2.41 to $\approx 0.50$ and
$\approx 0.47$, against an MPC-baseline scale of $\approx 0.38$. Bootstrap
CIs on these rates follow with the statistical pass
(\S\ref{sec:results}); the numbers above are single-pass point counts.

\subsubsection{Case studies (binding deliverable)}\label{sec:casestudies}

Four worked case studies, each carrying the four required elements (unsafe
proposal, authored physical-danger argument, gate verdict, triggering
constraint with signed margin), are drawn from the evaluation logs
(\code{data/case\_studies/c3\_gate\_eval\_2026-07-06/}): mid-drift
containment (cs01 --- the gate pins the follower drift just below the
bottoms bound that the ungated twin crosses on its way to $x_B = 0.69$);
blocked recovery (cs02 --- the agent's own return-to-spec is rejected as
negative-flow-infeasible and the ratcheted substitute is held instead);
spec-on-bound inadmissibility (cs03 --- the specification itself fails the
counterfactual at the best-behaved OP and the run is fatal by the
no-default-substitute rule); and the onset false negative (cs04 --- the
attractor's first ratifications pass with all nine margins positive: the
binding false-negative case). cs01/cs02 note the $y_D$-channel generality of
the same mechanism at the second follower OP.

\subsubsection{Cross-domain detector (resolved 2026-07-06: not run)}\label{sec:detector}

The timeboxed cross-domain detector experiment (the safety-layer descope, Appendix~\ref{app:records}) was not
started: its hard precondition --- a written feature-mapping memo resolving
the 52-variable TEP space onto the Column-A state space via
cross-plant-comparable derived features --- was never satisfied, and the
submission timeline bound on the gate evaluation and statistical
statistics instead. Learned early-warning layers on top of the
counterfactual gate, whether trained in-domain or transferred from public
benchmarks such as TEP, remain future work; the C3 contribution stands on
the rule-based counterfactual gate alone
(\S\ref{sec:gatemech}--\ref{sec:casestudies}), as the safety-layer descope provided.

\subsubsection{Scope boundary --- onset invisible, tail intercepted}\label{sec:scopeboundary}

The gate intercepts \emph{constraint-violating counterfactuals}, and the
evaluation shows precisely where that scope begins and ends against the
passive-follower attractor (\S\ref{sec:discussion}). The attractor's
\emph{onset} is invisible to the gate by construction: each early
ratification of the drifted plant state is a well-formed proposal whose
30-minute counterfactual is stable with all nine margins positive --- the
danger materializes only through the future sequence of targets the follower
policy will emit, which no single-proposal trajectory check evaluates (case
study cs04). The attractor's \emph{tail} is intercepted: once the drift has
progressed far enough that the forked trajectory leaves the pinned envelope
within the horizon, the gate blocks --- on the $x_B$ channel at one follower
OP and the $y_D$ channel at the other --- and contains the worst case by an
order of magnitude. The gate contains; it does not recover. Its
substitute action is the last \emph{accepted} target, which by interception
time has ratcheted with the waved-through onset, and the gate rejects the
agent's own late return-to-spec as dynamically infeasible (cs02). The
division of labor is therefore sharper than the pre-evaluation expectation
(which, extrapolating from the Nemotron screening trajectories, predicted no
interception at all): specification-anchoring is the supervisor's job --- a
prompt-level disambiguation eliminates the attractor at its source
(\S\ref{sec:discussion}) --- while the gate guarantees a bounded envelope
around whatever the supervisor does, including around its failures. Neither layer substitutes for the other. The boundary is a co-design
finding about gate and supervisor, not a gap to apologize for.

\section{Experimental Setup}\label{sec:setup}

\subsection{Disturbance scenarios, grids, seeds}\label{sec:scenarios}

The canonical scenario set at a given operating point comprises five arms:
feed-flow steps $F \pm 20\,\%$, feed-composition steps $z_F \pm 10\,\%$, and
a distillate-purity setpoint arm (\code{yD\_setpoint\_+0p5pct}). \textbf{As
executed by the evaluation driver, the set is four feed disturbances plus
one undisturbed control arm}: the setpoint arm's command is held at the
operator specification under the screening-pass canonical-target convention,
so no setpoint-tracking scenario is exercised (verified at code, cell, and
baseline-sweep level; disclosed as limitation \S\ref{sec:limitations} and
applied consistently wherever scenario coverage is claimed). Disturbances
step at $t = 5$~min on a 60-minute horizon (12 supervisory cycles at the
5-minute cadence).

Off-nominal evaluation uses the pre-registered 16-point operating grid
($F \in \{0.8, 0.9, 1.1, 1.2\}$ $\times$ $z_F \in \{0.45, 0.475, 0.525,
0.55\}$) with two pre-registered sub-metrics --- target acquisition
(off-nominal start, nominal MVs) and disturbance rejection (on-spec
pre-staged start) --- plus a 4-point screening grid used only for prompt
iteration. Stochastic configurations run $N = 10$ independent seeds per
cell; all statistics are distribution-level (no seed pairing --- same-seed
cross-run trajectories are not byte-stable at $T = 1.0$), reported as
medians and means with bootstrap 95\,\% confidence intervals (10\,000
resamples, seed-level resampling, fixed RNG seed) and Cohen's $d$ effect
sizes. Deterministic configurations (C0, C1) enter as point references.

\subsection{KPIs and statistical methodology}\label{sec:kpis}

\begin{figure}[tbp]
 \centering
 \includegraphics[width=\textwidth]{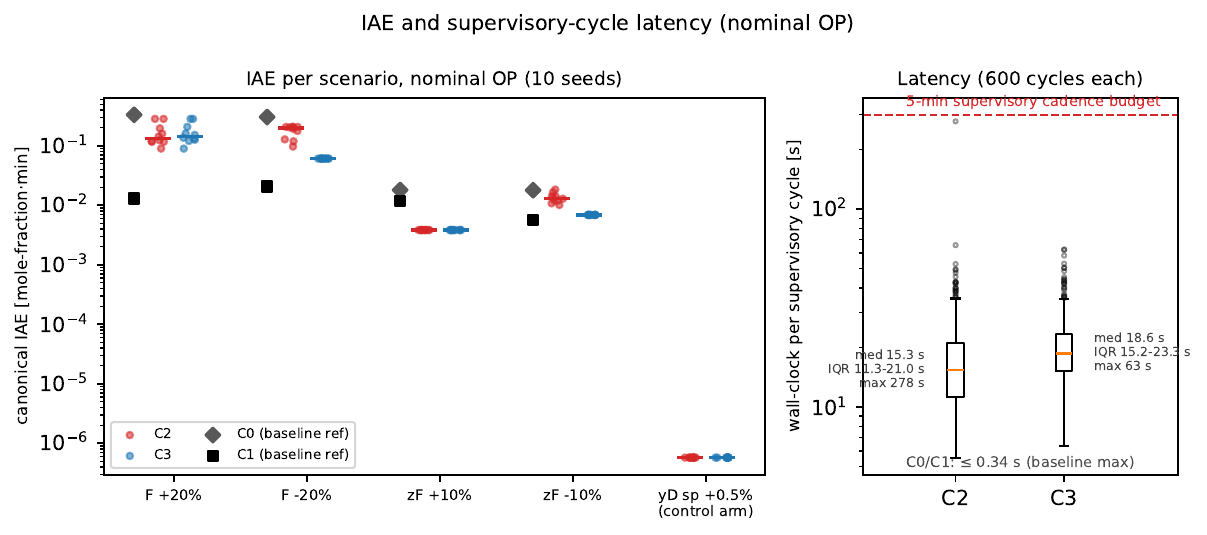}
 \caption{Per-scenario canonical IAE at the nominal OP (C2 companion and
 C3 statistical pass, 10 seeds each; C0/C1 as baseline point references; no
 C0/C1 markers on the undisturbed control arm) and supervisory-cycle
 latency (600 cycles per configuration) against the 5-min cadence budget
 with median / IQR / worst case annotated; the C0/C1 supervisory cost is
 sub-second and annotated, not drawn.}
 \label{fig:iaelatency}
\end{figure}

Five KPIs were pinned in the project KPI specification before the agent runs existed:

\begin{enumerate}
\item \code{aggregate\_iae} --- canonical IAE against the operator
 specification $(0.99, 0.01)$ over the scenario set, the primary
 performance KPI, with mandatory per-scenario disaggregation.
\item {\sloppy Off-nominal sub-metrics ---
 \code{off\_nominal\_\allowbreak target\_\allowbreak acquisition\_\allowbreak iae} and
 \code{off\_nominal\_\allowbreak disturbance\_\allowbreak rejection\_\allowbreak iae}, the outcome-class-B
 discriminators, evaluated against the Pareto-reference C1.\par}
\item {\sloppy Gate KPIs
 \code{constraint\_violation\_intercept\_rate} and
 \code{constraint\_violation\_detection\_rate} (outcome class~C), computed
 under the descope degeneracy protocol: the counterfactual \emph{is} the
 gate, so offline margin re-derivation audits log consistency rather than
 an independent detector ROC; both rates carry that caveat
 (\S\ref{sec:results}).\par}
\item \code{linearization\_consistency} --- the OP-locality of C1's
 linearization, reported as a structural property grounding the outcome-class-B
 narrative, not scored against.
\item \code{supervisory\_cycle\_wallclock} --- per-cycle decision latency
 against the deployability gate P95 $\leq 60$~s on the 5-minute cadence
 (Figure~\ref{fig:iaelatency}).
\end{enumerate}

Outcome classification follows a pre-registered decision tree over four
\textbf{outcome classes} (Appendix~\ref{app:records}): outcome class A (agent dominates on aggregate IAE), outcome class B (agent
wins an off-nominal sub-metric at $1.5\times$ against the Pareto reference),
outcome class C (gate is the load-bearing contribution: both gate rates $> 0.7$
with CI lower bounds clearing the threshold, plus $\geq 3$ documented case
studies), or Ambiguous --- an explicitly publishable landing zone. Every
threshold comparison requires bootstrap-CI separation, not point estimates.
The outcome classes, their thresholds, and the Methods paragraphs for each outcome
were \textbf{fixed before any C2 number existed}
(Appendix~\ref{app:records}); the headline classification itself is the record frozen 2026-07-02, prior to the statistical pass
(asymmetric outcome class B: target acquisition strong, disturbance rejection
inverse), quoted in \S\ref{sec:results}/\S\ref{sec:discussion} and never
recomputed.

\section{Results}\label{sec:results}

\subsection{Setup sanity: nominal ladder and C1 dominance on local regulation}
\label{sec:nominalsanity}

Two sanity checks come first. On the baseline-campaign nominal five-arm
scenario set, linear-MPC baseline C1 beats relay-tuned PID-only C0 on all
five scenario arms
(aggregate IAE margin $\approx 6.8\times$; frozen baseline record; the set
includes a setpoint arm that is not a feed disturbance---see
\S\ref{sec:limitations}), so the industrial baseline is strong rather than a
strawman. Under the same disturbance-rejection regime that C1 is designed for, the
ungated LLM supervisor is \emph{not} expected to win---and does not
(\S\ref{sec:bucket}). The off-nominal headline therefore asks a different question: where does
the adaptive supervisor add value once linearization stress and target
re-planning enter?

\subsection{Outcome classification (frozen V4-Flash headline)}\label{sec:bucket}

Table~\ref{tab:headline} freezes the 2026-07-02 classification on the
DeepSeek-V4-Flash 16-OP headline grid (before the statistical gate pass;
Appendix~\ref{app:records}). Cross-family results are scoped in
\S\ref{sec:crossfamily} and are not mixed here.

\begin{table}[tbp]
\centering
\caption{Frozen V4-Flash headline classification (C2/C1 P95-IAE ratio;
strong $\leq 0.50$, fails $> 0.67$). Point estimate in brackets. Full-grid
and agent-feasible aggregates both meet the same band.}
\label{tab:headline}
\small
\begin{tabular}{@{}llcc@{}}
\toprule
Sub-metric & Aggregate & Ratio (upper CI [pt]) & Band \\
\midrule
target acquisition & full 16-OP & $0.361$ [$0.351$] & strong \\
target acquisition & agent-feasible & $0.368$ [$0.359$] & strong \\
disturbance rejection & full 16-OP & $16.03$ [$10.18$] & fails \\
disturbance rejection & agent-feasible & $16.36$ [$10.45$] & fails \\
\bottomrule
\end{tabular}
\end{table}

On \code{target\_acquisition}, C2 beats the Pareto-reference C1 by
$2.77\times$ on the full grid (P95 IAE $58.55$ vs $162.08$) and $2.72\times$
on the agent-feasible regime that drops the LV-near-singularity cluster
($58.15$ vs $157.88$). Both land in the strong band (ratio $\leq 0.50$; the
looser inclusion threshold is $0.67$).

On \code{disturbance\_rejection} the same grid inverts: C1 beats C2 at an
upper-CI ratio of $16.03$ full-grid ($10.18$ at the point estimate; P95 IAE
$3.90$ vs $0.38$) and upper-CI $16.36$ agent-feasible ($10.45$ at the point;
P95 $4.01$ vs $0.38$; C1 is nearly invariant to the singularity exclusion).
Both land in the \code{fails} band.
Strong on re-planning, fails on local regulation: that split is the
headline.

\subsection{Failure geography and regime map}\label{sec:failuregeo}

Failures on the same 16-OP grid are spatially structured, not uniform
random attrition. On the completed Nemotron-3-Super cross-family
sweep (1600/1600 terminal; 686 done / 914 failed; all failures plant-side;
zero rate-limit/5xx), cell completion concentrates in the
\code{disturbance\_rejection} half (653/800 done across all 16 operating
points) while the \code{target\_acquisition} half largely collapses from
nominal manipulated-variable starts (33/800 done; only 8/16 OPs retain any
completed TA cell). Among feed-flow columns, $F{=}1.1$ is the hardest
(highest \code{exit\_2} share), and $F{=}0.8$ hosts the pre-registered
LV-near-singularity cluster --- both previously visible on V4-Flash. Figure~\ref{fig:regime} summarizes this regime map: LLM value and plant
operability co-locate with on-spec pre-staging and re-planning stress; local
disturbance rejection remains the MPC-favored regime.

\begin{figure}[tbp]
 \centering
 \resizebox{\textwidth}{!}{%
 \begin{tikzpicture}[
 font=\small,
 cell/.style={minimum width=1.7cm, minimum height=0.95cm, draw=black!70,
 align=center, inner sep=2pt},
 hdr/.style={minimum width=1.7cm, minimum height=0.5cm, align=center,
 font=\small\bfseries},
 ]
 \node[hdr] at (2.0,2.6) {$F{=}0.8$};
 \node[hdr] at (3.9,2.6) {$F{=}0.9$};
 \node[hdr] at (5.8,2.6) {$F{=}1.1$};
 \node[hdr] at (7.7,2.6) {$F{=}1.2$};
 \node[align=right, font=\small] at (0.15,1.6) {TA};
 \node[align=right, font=\small] at (0.15,0.5) {DR};
 \node[cell, fill=red!50] at (2.0,1.6) {low / LV};
 \node[cell, fill=red!48] at (3.9,1.6) {low};
 \node[cell, fill=red!58] at (5.8,1.6) {hardest};
 \node[cell, fill=red!42] at (7.7,1.6) {low};
 \node[cell, fill=green!38] at (2.0,0.5) {high};
 \node[cell, fill=green!40] at (3.9,0.5) {high};
 \node[cell, fill=green!32] at (5.8,0.5) {high};
 \node[cell, fill=green!38] at (7.7,0.5) {high};
 \end{tikzpicture}}
 \caption{Regime map (Super 1600-cell sweep, qualitative completion by
 $F$-column). \textbf{TA} row: target-acquisition half mostly plant-side
 collapse (only $33/800$ done overall). \textbf{DR} row: disturbance
 rejection mostly complete ($653/800$ done). Column done/failed (400 cells
 each): $F{=}0.8$ $184/216$; $F{=}0.9$ $155/245$; $F{=}1.1$ $195/205$
 (hardest, highest \code{exit\_2}); $F{=}1.2$ $152/248$. Headline KPI bands
 remain V4-Flash (TA strong $0.361$ upper CI; DR fails $16.03$ upper CI /
 $10.18$ point); Super supplies geography and a milder DR-fails
 \emph{point} band ($\approx 1.7$--$2.0$, not to be compared to V4's upper
 CI), with TA strong only survivorship-qualified
 (\S\ref{sec:crossfamily}).}
 \label{fig:regime}
\end{figure}

\subsection{Safety-gated agent at statistical scale --- experimental basis}\label{sec:c3basis}

C3 (agent + counterfactual gate, fixed v1 prompt, DeepSeek-V4-Flash)
is evaluated on four operating points --- nominal $(1.0, 0.5)$, the
follower-attractor OP $(1.2, 0.55)$, and two control OPs $(1.2, 0.45)$ and
$(0.8, 0.55)$ --- crossed with the canonical five-scenario set and ten seeds
(250 cells including the C2-nominal companion baseline). As executed by the
evaluation driver, the canonical set comprises \textbf{four feed
disturbances ($F \pm 20\,\%$, $z_F \pm 10\,\%$) plus one undisturbed control
arm}: the \code{yD\_setpoint\_+0p5pct} scenario's setpoint command is held
at the operator specification by the screening-pass convention, so no
disturbance reaches the plant, the agent, or the KPI on that arm (verified
at code, cell, and baseline-sweep level). This wording is
used consistently with the Methods scenario description. All C3-vs-C2
comparisons are distribution-level over seeds (no seed pairing; same-seed
cross-run trajectories are not byte-stable at $T = 1.0$); we report medians
with bootstrap 95\,\% CIs (10\,000 resamples) and Cohen's $d$. One attractor
cell (the cold-start attractor-OP fatality, below) terminates at cycle 0 and contributes no IAE; its
stratum aggregates $n = 9$ without imputation.

\subsection{Containment at the follower operating points}\label{sec:containment}

\begin{figure}[tbp]
 \centering
 \includegraphics[width=\textwidth]{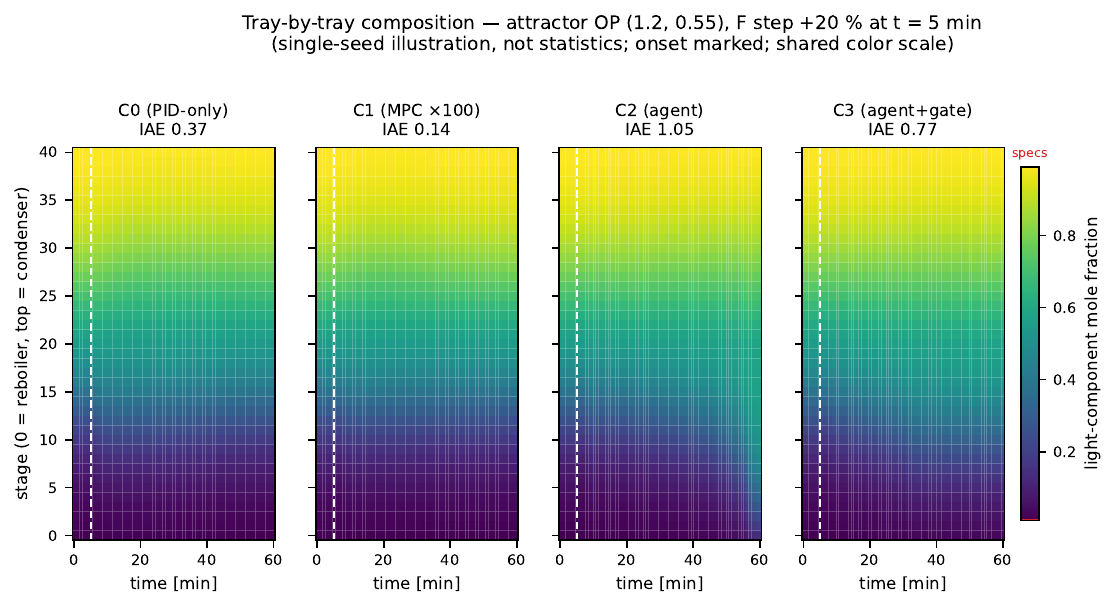}
 \caption{Tray-by-tray composition over time, one panel per configuration,
 at the attractor OP $(1.2, 0.55)$ under \code{F\_step\_+20pct} (onset
 marked; shared color scale; per-panel headline IAE). \textbf{Single-seed
 illustration, not statistics}; population-level comparisons are given in
 Figures~\ref{fig:iaelatency} and~\ref{fig:peakdev} and
 \S\ref{sec:containment}.}
 \label{fig:tray}
\end{figure}

\begin{figure}[tbp]
 \centering
 \includegraphics[width=0.92\textwidth]{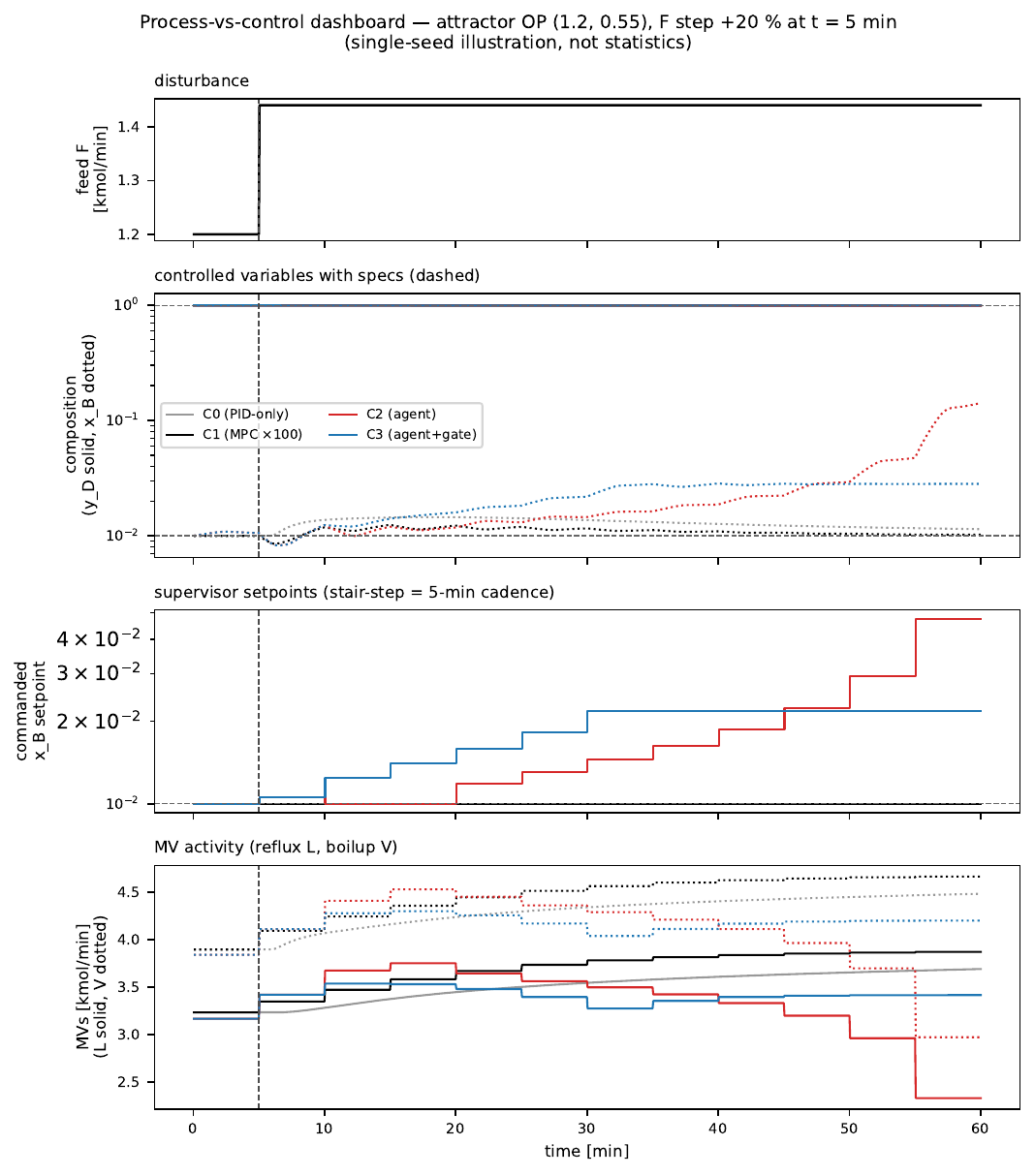}
 \caption{Process-vs-control dashboard for the same cell set as
 Figure~\ref{fig:tray}: disturbance profile, controlled variables with spec
 bands, commanded setpoint staircases (5-min supervisory cadence), and MV
 activity (reflux $L$, boilup $V$). \textbf{Single-seed illustration, not
 statistics}; population-level comparisons are given in
 Figures~\ref{fig:iaelatency} and~\ref{fig:peakdev} and
 \S\ref{sec:containment}.}
 \label{fig:dashboard}
\end{figure}

\begin{figure}[tbp]
 \centering
 \includegraphics[width=\textwidth]{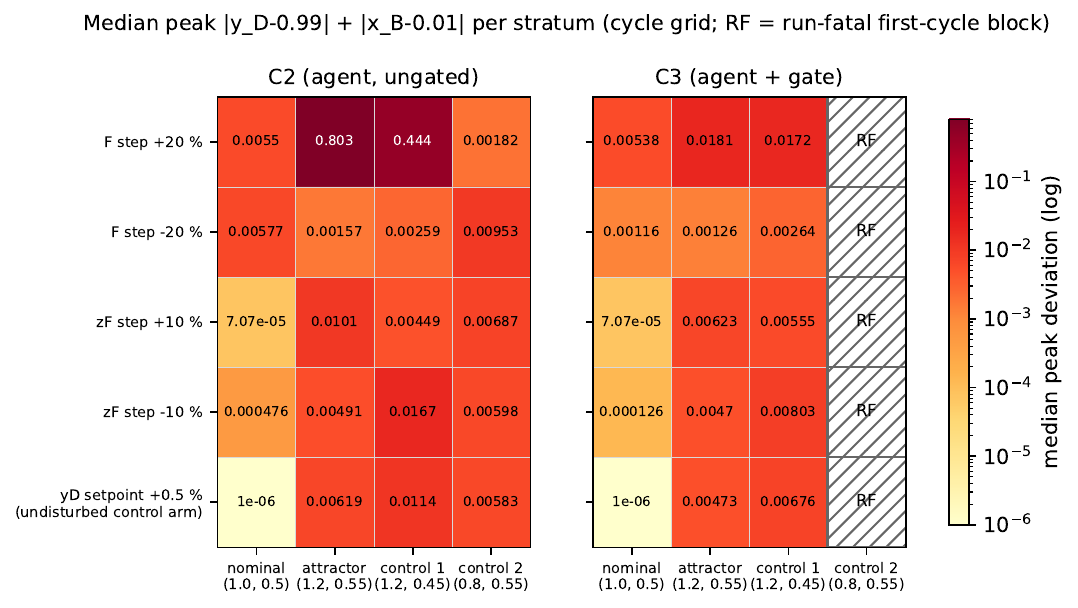}
 \caption{Median peak post-onset composition deviation
 $|y_D - 0.99| + |x_B - 0.01|$ per stratum, C2 vs C3, all four OPs
 $\times$ the five-arm scenario set (log color scale; RF = run-fatal
 first-cycle block; the cold-start attractor-OP fatality excluded per the results-memo rule, its
 stratum aggregating $n = 9$). Peaks are sampled on the 5-min supervisory
 grid and underestimate intra-cycle peaks (measured example
 $\approx 2.7\times$); the C2-vs-C3 contrast is resolution-consistent
 because both configurations are sampled identically, and no absolute
 comparison against the full-resolution C0/C1 trajectories is drawn from
 this figure.}
 \label{fig:peakdev}
\end{figure}

The gate's containment of the passive-follower failure mode, first observed
in the 40-cell gate evaluation, is \textbf{replicated by an independent
re-run of the same two strata at equal $n$} (10 fresh seeds per stratum) on
the scenario where the follower occurs (\code{F\_step\_+20pct}); what this
pass adds beyond the replication is scenario breadth (all five arms at every
OP). At the attractor OP the C3 canonical-IAE median is 0.486 [0.401, 0.642]
against a C2 baseline of 6.401 [0.282, 11.464] (Cohen's $d = -1.44$); the
P95 compresses from 11.464 to 0.774. At control 1 the $y_D$-channel follower
is contained from a C2 median of 2.409 [0.273, 9.483] to 0.514
[0.320, 0.575] ($d = -1.23$; P95 $9.483 \rightarrow 0.604$). The C2 baseline
distributions are bimodal --- 6/10 seeds enter the follower regime at the
attractor and 5/10 at control 1 (headline-sweep diagnostics) --- and the
width of the C2 median CIs ([0.282, 11.464], [0.273, 9.483]) reflects this
follower incidence rather than smooth dispersion. The trigger channels match
the mechanism: 23/120 blocks at the attractor, all \code{x\_B\_lower\_0.03};
24/120 at control 1, all \code{y\_D\_lower\_0.97}; all fast-fail. C3 lands
near --- but not at --- the C1 reference level for this sub-metric
(grid-scale disturbance-rejection IAE $\approx 0.38$, cited from the
frozen classification record, Appendix~\ref{app:records}). On the remaining disturbance scenarios at these OPs the
C3 and C2 IAE distributions are statistically close ($|d| \leq 0.56$,
overlapping CIs; Figure~\ref{fig:peakdev} gives the peak-deviation view of
the full matrix). Figures~\ref{fig:tray} and~\ref{fig:dashboard} illustrate
the follower-vs-contained mechanism on a single representative seed at the
attractor OP (single-seed illustration, not statistics).

Because intra-cycle trajectories are not persisted by the evaluation driver,
time-to-recovery is reported as a censoring statement rather than a figure
channel: \textbf{at a $10^{-3}$ recovery band, no probed cell of either
configuration recovers within the 60-min horizon} --- recovery-time
comparisons are therefore not resolvable on this dataset.

\subsection{The spec-on-bound intercept class and the inoperable control OP}\label{sec:specbound}

The largest intercept class of the pass is not the follower containment but
a second class on the $y_D$-raising disturbances (\code{F\_step\_-20pct},
\code{zF\_step\_+10pct}) and --- unexpectedly --- on
\code{zF\_step\_-10pct}: uniformly \code{y\_D\_upper\_0.99}, 46--110 blocks
per affected stratum. Mechanistically, the regulatory
MPC's rejection of either $z_F$ step or the $F - 20\,\%$ step settles $y_D$
marginally \emph{above} the pinned 0.99 bound ($+3.6\times 10^{-5}$ to
$+6.3\times 10^{-4}$, i.e.\ 36--630$\times$ the numerical tolerance
$\varepsilon$), and because the nominal specification sits exactly on that
bound, the counterfactual fork starts --- or immediately lands --- in
violation. This falsifies pre-registered expectation 3 (``interception
concentrated on \code{F\_step\_+20pct}, other scenarios mostly
pass-through''): the spec-on-bound mechanism, not the follower attractor,
dominates intercept volume across the grid.

At control 2 $(0.8, 0.55)$ the same mechanism is total: \textbf{all 50 cells
are run-fatal at cycle 0}, exactly as pre-registered (expectation 1). The 50
intercept rows are byte-uniform across scenarios and seeds --- the cycle-0
proposal is the specification itself, the trigger is
\code{y\_D\_upper\_0.99}, the first fork violation occurs at $t = 0.1$~min
at a margin of $-4.83\times 10^{-4}$ (483$\times \varepsilon$). Since the
fork fails before the disturbance onset ($t = 5$~min), the
scenario-independence holds by construction: what the gate rejects is the
regulatory layer's own pre-disturbance spec-tracking transient at this OP. A
zero-margin specification plus a trajectory gate thus renders the
\emph{well-behaved} low-$F$ operating point inoperable while the misbehaving
OPs are merely contained.

One additional first-cycle fatality occurred outside control 2:
\textbf{the cold-start attractor-OP fatality} (attractor $\times$ \code{yD\_setpoint\_\allowbreak +0p5pct}, seed 2)
--- notably on the undisturbed control arm. The agent proposed
$(0.98, 0.02)$ at cycle 0 of an undisturbed hold; the fork tripped
\code{x\_B\_lower\_0.03} and, with no previously accepted target available,
the run terminated. This punctually falsifies pre-registered expectation 4
(``no first-cycle blocks at the two $F = 1.2$ OPs'') and shows the run-fatal
admission path being exercised with no disturbance present.

\subsection{The nominal point: what the gate costs, and what it does instead}\label{sec:nominal}

The companion run gives the nominal C2 baseline, and the pre-registered
companion expectation --- ``C2-nominal and C3-nominal distributions
indistinguishable; the gate costs nothing at the nominal point (0
intercepts, IAE distributions congruent)'' --- is scored per scenario:
\textbf{confirmed on 2 of 5} (\code{F\_step\_+20pct}: 0/120 intercepts,
$d = +0.09$; \code{yD\_setpoint\_+0p5pct}: 0/120, both configurations at IAE
$\approx 0$ on the undisturbed arm, trivially congruent) and
\textbf{falsified on 3 of 5} (\code{F\_step\_-20pct},
\code{zF\_step\_+10pct}, \code{zF\_step\_-10pct}: 110/120 intercepts each).
The falsification runs in an unexpected direction: where the gate
intervenes, C3's IAE is equal to or \emph{lower} than C2's ($F - 20\,\%$:
0.061 vs 0.198; $z_F - 10\,\%$: 0.0069 vs 0.013; $z_F + 10\,\%$: identical
0.0038 in both configurations on every seed). In these gate-dominated strata
the C3 seed variance is degenerate (bit-identical IAE across seeds; see
below), so the pooled standard deviation of any effect size is carried
entirely by the C2 side; the Cohen's $d$ values there ($-3.73$ for
$F - 20\,\%$, $-3.66$ for $z_F - 10\,\%$) are therefore reported as
\textbf{descriptive magnitudes only} --- the medians with their disjoint
bootstrap CIs carry the comparison. The cost claim survives on IAE while the
mechanism claim (zero intercepts) fails.

The nominal strata form a clean trichotomy. On the
free arms (\code{F\_step\_\allowbreak +20pct}, \code{yD\_setpoint\_\allowbreak +0p5pct}) the gate
passes everything and the agent's proposals execute. On the three
gate-dominated arms every seed runs at 11/12 blocks and the cells are fully
\textbf{agent-decoupled}: cycle 0 executes the specification, every later
proposal is blocked and substituted with the specification, and the closed
loop is consequently LLM-independent from cycle 1 --- which is why the C3
IAE is bit-identical across seeds in these strata. Across the whole pass the
590 blocks split into \textbf{318 operationally effective} (an off-spec
proposal replaced by the specification --- the substitutions that produce
the C3 $<$ C2 nominal medians), \textbf{221 no-ops} (the specification
blocked and substituted with itself; 110/110 of the \code{zF\_step\_+10pct}
blocks), and \textbf{51 run-fatal without substitute} (50 at control 2, plus
the cold-start attractor-OP fatality).

\subsection{Gate rates under the degeneracy protocol}\label{sec:gaterates}

Re-deriving the ground-truth label offline from the raw logged margins
(unsafe $\Leftrightarrow$ any margin $< -\varepsilon$), independently of the
stored block flag, over all 1\,839 gated proposals yields zero mismatches
and zero counterfactual integration failures; hence
\code{constraint\_violation\_intercept\_rate} $= 1.000$ (bootstrap 95\,\% CI
[1.000, 1.000]) and \code{constraint\_violation\_detection\_rate} $= 1$.
These are reported strictly under the evaluation protocol's caveat, quoted
verbatim: ``the pinned \S3.2 definitions predate the safety-layer descope (Appendix~\ref{app:records}) and assume a
detector that blocks and a counterfactual that verifies. With the
counterfactual \emph{as} the gate, the ground-truth label
\code{counterfactual\_unsafe\_p} is re-derived \textbf{offline from the raw
logged margins} (unsafe $\Leftrightarrow$ any margin $< -\varepsilon$),
independent of the stored \code{blocked} flag --- an audit of log
consistency rather than an independent oracle. Consequences, reported as
such: intercept\_rate $= 1 -$ (share of
\code{counterfactual\_integration\_failure} blocks, which are conservative
blocks without a confirmed violation); detection\_rate $= 1$ by construction
within this ground truth. The \emph{substantive} false-negative surface ---
sequence-level unsafety of onset ratifications (cs04) --- lies outside this
formal ground truth by construction and is carried by the case studies; the
paper reports both rates with this caveat verbatim rather than presenting
the degenerate 1.0 as an empirical victory.'' Concretely, the cs04 onset
surface is non-empty in this pass: the attractor's early-drift proposals
pass the gate on every seed before the containment blocks begin.

\subsection{Pre-registered expectations --- scorecard}\label{sec:scorecard}

Table~\ref{tab:scorecard} scores pre-registered expectations, including
falsifications. Several misses are \emph{geometry}: blocked forks still
violate the pinned bound by $36$--$630\times\varepsilon$.

\begin{table}[tbp]
\centering
\caption{Pre-registered expectation scorecard (statistical gate pass).}
\label{tab:scorecard}
\footnotesize
\setlength{\tabcolsep}{3.5pt}
\begin{tabular}{@{}p{5.8cm}ll@{}}
\toprule
Expectation & Verdict & Note \\
\midrule
Control 2 run-fatal, all 50 cells & confirmed & 50/50; \S\ref{sec:specbound} \\
Nominal $\approx 0$ intercepts & falsified (3/5 arms) & geometry, not noise \\
Intercepts concentrated on $F{+}20\%$ & falsified & spec-on-bound dominates \\
No first-cycle blocks at $F{=}1.2$ & punctually falsified & cold-start attractor arm \\
Gate costs nothing at nominal & 2/5 conf., 3/5 fals. & C3 $\leq$ C2 IAE when fals. \\
Direction hypothesis ($y_D$ raise/lower) & both arms falsified & pre-outcome in run-spec \\
\bottomrule
\end{tabular}
\end{table}

\section{Discussion}\label{sec:discussion}

\subsection{Asymmetric division of labor}\label{sec:asymmetry}

Table~\ref{tab:headline} is the finding: the comparison \emph{splits} by
sub-metric. Aggregate ``agent better/worse than MPC'' is the wrong frame.

On target acquisition the strong band fits the prior that an LLM supervisor
can re-plan multi-step trajectories when C1's per-OP linearization is
stressed. The QP is then locally inconsistent with the plant; C2 can
retarget. Full-grid and agent-feasible aggregates both meet ratio
$\leq 0.50$, so the band does not depend on dropping the LV-near-singularity
cluster.

On disturbance rejection the inverse is first-class evidence, not a caveat.
Steady-state regulation at a known operating point is where LLM supervisors
do not belong: fixed-weight linear MPC beats the agent by an order of
magnitude. Failures on the headline grid are spatially narrow (one boundary
cluster at $(F{=}1.1,z_F{=}0.55)\times\code{F\_step\_+20pct}$ at exactly
$5/10$, below the strict $>50\,\%$ high-failure rule). That locus echoes the
screening cluster that produced the passive-follower attractor: seeds that
read a disturbed plant as a new specification rather than holding the
operator target. The attractor is prompt ambiguity, not proof of a hard
capacity bound (\S\ref{sec:optiona}).

Guidance is simple: agent for re-planning under linearization stress; MPC
for local regulation where the linearization holds. The gate discussion
(\S\ref{sec:gatediscussion}) makes that split operational.

\subsection{Prompt-disambiguation sensitivity}\label{sec:optiona}

\emph{(Sensitivity evidence, never replacement headline material --- all
\S\ref{sec:results} headline classifications remain those of the fixed v1
prompt.)}

The \S\ref{sec:asymmetry} disturbance\_rejection discussion attributes the
passive-follower failure mode to a specification ambiguity in the fixed v1
system prompt: ``minimize IAE'' admits both a hold-the-specification and a
track-the-plant reading, and the canonical IAE definition --- which makes
hold-the-specification the correct reading --- is not visible to the agent.
The interpretation was tested directly. Variant \code{v2\_disambiguated}
appends one disambiguating instruction to the fixed v1 prompt and changes
nothing else (additive structure machine-checked in the test suite). Thirty
cells---the attractor OP plus two controls, ten seeds each, same scenario,
sampling, and seeds as the headline run---ran on the same DeepSeek-V4-Flash /
Novita stack.

\begin{table}[tbp]
 \centering
 \caption{Passive-follower incidence and cell IAE under v1 (headline) vs
 \code{v2\_disambiguated} (sensitivity), \code{F\_step\_+20pct} $\times$
 disturbance\_rejection, DeepSeek-V4-Flash.}
 \label{tab:optiona}
 \footnotesize
 \setlength{\tabcolsep}{3.5pt}
 \begin{tabular}{llllll}
 \toprule
 OP $(F, z_F)$ & Role & v1 incid. & v1 med [max] IAE & v2 incid. & v2 med [max] IAE \\
 \midrule
 $(1.2, 0.55)$ & attractor cell ($x_B$ channel) & 6/10 & 6.40 [11.46] & \textbf{0/10} & 0.126 [0.126] \\
 $(1.2, 0.45)$ & control 1 ($y_D$ channel, see note) & 5/10 & 2.41 [9.48] & \textbf{0/10} & 0.100 [0.100] \\
 $(0.8, 0.55)$ & control 2 (clean) & 0/10 & 0.079 [0.120] & 0/10 & 0.037 [0.047] \\
 \bottomrule
 \end{tabular}
\end{table}

At the attractor cell the incidence drops from 6/10 to 0/10 (two-proportion
bootstrap 95\,\% CI of the drop [0.30, 0.90]; Fisher exact $p = 0.011$); all
ten v2 seeds hold the specification for all twelve cycles. The clean control
shows no regression --- the disambiguation does not disturb the
correct-behavior regime.

\paragraph{Note on control 1 (readout-time criterion extension, disclosed).}
The base readout protocol pre-registered the attractor signature on the
$x_B$ channel, per the original Nemotron forensic. During the sensitivity
readout, the v1 baseline at control 1 --- expected well-behaved from the
Nemotron reference --- was found to carry the same follow-the-plant
mechanism on the \emph{$y_D$ channel} (five of ten seeds walk
\code{y\_D\_target} down with the sagging distillate purity until the column
collapses; the agent's rationale states the mechanism verbatim). The
detection criterion was therefore generalized to the symmetric channel
\emph{at readout time}, and applied identically to v1 and v2 cells; it reads
only the target-vs-output trajectories, never the IAE that enters any
verdict. Under v2 this channel's incidence is also 0/10. No headline
aggregate is affected: the control-1 cells enter the frozen classification's tables
unchanged, and this classification exists only in this sensitivity analysis.

Two secondary observations. First, within every v2 cell all ten seeds
produce numerically identical trajectories: removing the interpretation
ambiguity also removes the sampling lottery that made the v1 failure
deterministic-per-seed, which supports the reading that the attractor is an
interpretation-selection phenomenon rather than a capacity limit. Second,
the mechanism is channel-general and plant-state-dependent ($x_B$ channel
where the bottom composition drifts, $y_D$ channel where the distillate
sags), which is consistent with a specification-anchoring failure and not
with a channel-specific control weakness.

\paragraph{Scope and future work.}
The sensitivity result is established on DeepSeek-V4-Flash, the model of the
headline evaluation (Appendix~\ref{app:records}); the attractor itself
was first observed on Nemotron-120B (3-corner screening) and reproduces
across model family, provider, and sampling temperature from the same v1
prompt. A v2 verification on the Nemotron side was not performed --- the NIM
serving capacity available to this study did not permit further sweeps ---
and remains future work; until then, the elimination result is reported as
model-conditional on V4-Flash. The v1 $\rightarrow$ v2 comparison is a
sensitivity analysis by design: all headline classifications in
\S\ref{sec:results} remain those of the fixed v1 prompt under the
pre-registered no-retroactive-prompt-modification commitment (Appendix~\ref{app:records}).

\subsection{Safety gate --- containment, spec-on-bound geometry, and what
the gate is for}\label{sec:gatediscussion}

\paragraph{Containment without recovery --- now with a quantitative recovery
bound.}
At both follower OPs the gate converts an unbounded
specification-abandonment failure into a bounded off-spec offset, and the
statistical-pass independent re-run at equal $n$ replicates the 40-cell result:
worst-case cell IAE falls by an order of magnitude (P95
$11.46 \rightarrow 0.774$; $9.48 \rightarrow 0.604$), medians from 6.40 and
2.41 to 0.486 and 0.514, against an MPC-baseline scale of $\approx 0.38$
(Appendix~\ref{app:records}). What the gate cannot do is restore the
specification: its substitute is the last accepted target, which has
ratcheted with the onset drift it correctly (by its own contract) admitted.
The statistical pass makes ``containment, not recovery'' quantitative rather
than anecdotal: \textbf{at a $10^{-3}$ recovery band, not a single probed
cell of either configuration returns to the specification within the
60-minute horizon} --- recovery is not merely imperfect, it is unobserved at
this horizon.

\paragraph{Scoping note on the blocked-recovery case study (cs02).}
The 2026-07-06 evaluation additionally observed the gate rejecting the
agent's own return-to-spec proposals (7 of 48 follower-OP blocks). This
phenomenon \textbf{did not recur in the statistical pass (0 of 47 blocks in the
same strata)}, and the 2026-07-06 singleton triggers (\code{negative\_flow},
\code{x\_B\_upper}) are likewise absent --- consistent with non-byte-stable
$T = 1.0$ sampling across runs. Blocked recovery is therefore a property of
individual trajectories, not a stable population feature at $n = 10$; case
study cs02 stands on the 2026-07-06 artifacts and is scoped as such.

\paragraph{Spec-on-bound admissibility --- from single-OP finding to the
dominant intercept class.}
The 2026-07-06 evaluation found one operating point (control 2) where the
specification itself fails the counterfactual. The statistical-pass scenario grid
shows this was not an outlier but the \textbf{dominant intercept class of
the entire pass}: 590 blocks were recorded, and \textbf{534 of them} carry
the \code{y\_D\_upper\_0.99} trigger produced by one geometry --- the
nominal specification sits exactly on the pinned purity bound, and the
regulatory MPC's rejection of a $y_D$-raising (or, unexpectedly,
$y_D$-lowering) feed disturbance settles $y_D$ marginally \emph{above} that
bound ($+3.6\times 10^{-5}$ to $+6.3\times 10^{-4}$, i.e.\ 36--630$\times$
the numerical tolerance). At the nominal operating point three of the five
scenario arms run gate-dominated at 11/12 blocks on every seed. Control 2 is
the extreme case of the same geometry, not a separate phenomenon: there the
regulatory layer's \emph{pre-disturbance} spec-tracking transient already
overshoots the bound ($-4.83\times 10^{-4}$ at $t = 0.1$~min on the fork,
byte-uniform across all 50 cells and all five scenarios), so no first
proposal is admissible and every run is fatal under the
no-default-substitute rule --- the disturbance never has a chance to matter.
Gate admissibility is governed by the regulatory layer's settling/overshoot
profile at the operating point, not by how well-behaved the supervisor is. A
post-hoc single-trajectory illustration attributes the
offset at the MV level: the stationary regulatory correction to a
$y_D$-raising disturbance is almost pure boilup ($\Delta V \approx
+0.043$~kmol/min against $\Delta L \approx -0.008$ at settling), which holds
$x_B$ just under specification while the un-cancelled vapor surplus carries
$y_D$ tens of $\varepsilon$ above the pinned bound --- the gate prices
exactly this trade. The same illustration's C0 contrast shows the identical
MV signature without any MPC in the loop: the offset is a property of
regulatory-layer disturbance rejection under spec targets at this operating
point, not of the particular regulatory controller.

\paragraph{The nominal point: a trichotomy, and an ambivalent companion
result.}
The nominal strata split three ways: two free arms (0/120 intercepts each),
and three gate-dominated arms in which the cells are fully
\textbf{agent-decoupled} --- cycle 0 executes the specification, every later
proposal is blocked and substituted with the specification, and the closed
loop is LLM-independent from cycle 1 (bit-identical IAE across seeds). The
pre-registered companion expectation (``the gate costs nothing at the
nominal point: 0 intercepts, congruent IAE distributions'') was confirmed on
2 of 5 arms and falsified on 3 of 5 --- but falsified in a direction the
expectation did not contemplate: wherever the gate intervened, C3's IAE was
equal to or \emph{lower} than C2's. The result is double-edged: the gate restores
baseline disturbance-rejection behavior \textbf{because the v1 agent's
proposals at the nominal point actively hurt} --- 318 of the 590 blocks
across the pass replaced an off-spec agent proposal with the specification
(operationally effective corrections), against 221 no-op blocks (the
specification blocked and substituted with itself) and 51 run-fatal
first-cycle blocks. Read as gate evidence, the nominal result says the gate
is a competent specification-holder; read as agent evidence, it says that at
the nominal point the supervisory layer under the v1 prompt adds negative
value that the gate must remove. Both readings are correct, and the second
is the \S\ref{sec:asymmetry}/\S\ref{sec:optiona} disturbance\_rejection
finding reappearing inside the C3 configuration.

\paragraph{A proposal-quality finding without any disturbance.}
One first-cycle fatality occurred outside control 2: at the attractor OP, on
the \emph{undisturbed control arm} of the scenario set, one seed's cycle-0
proposal $(0.98, 0.02)$ failed the counterfactual (\code{x\_B\_lower\_0.03})
with no previously accepted target to fall back on. This is a
proposal-quality observation in its own right: the run-fatal admission path
can be exercised by the agent alone, at a steady operating point, with no
disturbance present --- the gate's first-cycle admission rule is exposed to
the supervisor's cold-start proposal distribution, not only to plant
excursions.

\paragraph{A falsified direction hypothesis, reported as a methodological
strength.}
Mid-run, before the corresponding cells executed, we registered a direction
hypothesis for the nominal intercepts ($y_D$-raising disturbances collide
with the spec-on-bound limit; $y_D$-lowering disturbances stay clean; the
setpoint arm blocks) --- documented with timestamps in the run-spec
amendment. \textbf{Both predictive arms were falsified}: the
$z_F - 10\,\%$ arm blocks 11/12 on every seed (the closed-loop $y_D$ settles
above the bound regardless of the disturbance sign --- the hypothesis'
premise about the transient direction was wrong), and the setpoint arm
passes untouched (mechanically: the arm is an undisturbed hold under the
evaluation convention, \S\ref{sec:limitations}). Under the project's
falsifiable-claims positioning (Appendix~\ref{app:records}) this is reported as evidence the
pre-registration machinery works as intended: the registered prediction was
specific enough to be wrong, the falsification was detected by counting, and
the corrected mechanism (a sign-independent positive settling offset of the
regulatory layer at a spec-on-bound operating point) is better-supported
than the hypothesis it replaced.

\paragraph{Design implication (sketched only; deliberately not implemented
in this study).}
The co-design requirement stands --- specification, constraint envelope, and
admission rule must be designed together --- but the statistical-pass data sharpen
the lever priority. \textbf{Spec back-off} (operating targets strictly
interior to the safety envelope, $\delta$ sized to the regulatory settling
offset) is now clearly the first lever: it addresses the dominant intercept
class in one move --- the 330 nominal blocks, the 50 control-2 fatalities,
and the gate-dominated agent-decoupling all trace to the zero-margin
specification, and the measured settling offsets ($3.6\times 10^{-5}$ to
$6.3\times 10^{-4}$, plus the $4.8\times 10^{-4}$ control-2 transient) give
$\delta$ an empirical scale. \textbf{Overshoot-aware first-cycle admission}
(transients judged against a settling-window rather than the full horizon)
is the second lever, targeting the two run-fatal paths (control 2; the
no-disturbance cold-start fatality). A \textbf{recovery-capable substitute
policy} drops to third: only the follower tails need it, and the
blocked-recovery interaction it would fix did not even recur in the statistical
pass. Evaluating the levers still requires re-running the evaluation matrix
per lever and remains future work --- the present contribution is the
measured demonstration of which levers are load-bearing, and in what order.

\subsection{Cross-family scoping}\label{sec:crossfamily}

\emph{(Cross-family paragraphs reproduced from the classification record frozen 2026-07-02, prior to the statistical pass; model-conditional per the same resolution --- V4-Flash-only scoping; terminology normalized to this paper's usage; all numerical values verbatim per the frozen record; original wording preserved in the repository decision record, Appendix~\ref{app:records}.)}

\begin{quote}
``On the V4-Flash 16-OP headline grid two architectural difficulty zones
surface in the Skogestad LV configuration: the $F{=}0.8$ low-$z_F$ cluster
where the LV linearization becomes near-singular, and the $F{=}1.1$
high-load column where the per-cluster failure rate is highest across the
four $F$ columns. Both zones manifest primarily as
\code{target\_acquisition}-from-nominal-MVs failures, and both belong to the
plant/baseline geometry --- the LV configuration's nonlinearity is hard for
any controller paired with a Pareto-tuned linear MPC, independent of which
supervisory layer sits above it. This statement is supported by the V4-Flash
data alone; it characterizes the control problem, not the supervisory
agent.''
\end{quote}

\begin{quote}
``The asymmetric outcome-class-B signature itself (target\_\allowbreak acquisition strong +
disturbance\_\allowbreak rejection inverse) is, on the 16-OP headline grid, a V4-Flash
observation. A preliminary cross-model hint is provided by the Nemotron-120B
three-corner screening (documented separately), which exhibits the
same asymmetric band structure (target strong, dist fails) at the three
overlapping corners; the magnitude differences are expected given the
different grids and different model families. The cross-provider context
(Nemotron on NIM-NVFP4, V4-Flash on Novita-fp8) already supports an
open-weights, multi-provider reproducibility chain for the asymmetric band
structure, independent of the headline outcome.''
\end{quote}

\textbf{Cross-family result (2026-07-24, updating the earlier
unresolved-by-data closure).} The cross-family test was re-instantiated as a
full 1600-cell Nemotron-3-Super-120B sweep on a throttle-free paid endpoint
(DeepInfra bf16), at the per-model DoE-pinned operating point ($T = 0.3$)
that mirrors the V4-Flash headline's own DoE pin ($T = 1.0$) --- each model
at its validated sampling point rather than a shared temperature. The run
completed cleanly (686 done / 914 failed, all failures plant-side, zero
infrastructure rate-limits; numbers verbatim per the frozen record, not recomputed). The outcome is
threefold and stated without escalation:
\begin{itemize}
\item \textbf{Disturbance-rejection arm --- band direction replicates.}
Super's disturbance-rejection arm is well covered (653/800 done cells,
all 16 operating points) and falls in the same \code{fails} band as
V4-Flash (Super C2/C1 $\approx 1.69$ full grid, $1.95$ agent-feasible); the
\emph{direction} of the inverse asymmetry reproduces on a second family, but
its \emph{magnitude} is model-conditional: Super point ratios
$\approx 1.7$--$2.0$ versus V4-Flash \emph{point} $\approx 10.2$ (upper CI
$16.03$ in Table~\ref{tab:headline})---compare point-to-point or CI-to-CI,
never Super-point to V4-upper-CI. Replication is claimed at the band level
only.
\item \textbf{Target-acquisition arm --- survivorship-qualified.} Super's
\code{target\_acquisition} half collapsed plant-side (33/800 done cells,
8/16 operating points). The surviving cells' \code{strong} band
($\approx 0.335$/$0.344$) is numerically close to V4-Flash but rests on a
collapse-survivor subsample and is \emph{not} read as a cross-family
confirmation of the target-acquisition band.
\item \textbf{Protocol operability is itself model-conditional.} Whether a
family stays on the pinned JSON contract in the hard target-acquisition
zones differs by model: Super collapses that half 767/800 against
$\approx 416/800$ for V4-Flash; Nemotron-3-Ultra was not protocol-capable
(0/20) and GLM-5.2 was not protocol-capable (0/10). Operability/attrition is
a cross-family property in its own right.
\end{itemize}
Three axes are confounded by design: model family; per-model DoE sampling
($T = 0.3$ Super vs $T = 1.0$ V4-Flash); and \textbf{reasoning realization}
(Super first-round realized-off vs V4-Flash Novita realized-on;
\S\ref{sec:llmspec}). The Super-vs-Ultra capacity axis is separately
confounded. No pure family effect is isolated.
The 2026-07-15 sweep is now the \emph{primary} cross-family dataset; the
earlier Nemotron NIM partial (500 done cells) remains scope-limited
supplementary evidence (Appendix~\ref{app:supplementary}). The \textbf{frozen
V4-Flash classification is untouched and remains this paper's headline}; the
Super result feeds this section and the Supplementary only. All
\S\ref{sec:results} and \S\ref{sec:discussion} headline claims stay
model-conditional on DeepSeek-V4-Flash unless explicitly marked otherwise.

\subsection{MPC-free deployment economics}\label{sec:mpcfree}

Beyond the primary ladder, we ask a deployment-economics question: can an
operator skip the MPC layer entirely --- no license, no commissioning, no
model maintenance, the practice burden documented in the industrial-MPC
literature \cite{darby2012mpc} --- and run the agentic supervisor directly
over the existing PID regulatory layer? The architecture makes this a
configuration flag rather than a redesign: the agent is backend-agnostic,
and the PID backend is the unmodified C0 regulatory stack with no retuning
(deliberately --- the deployment premise is that the operator removes the
MPC and changes nothing else).

This branch is evaluated at \textbf{demonstration scope}: a single
scenario-set comparison (the canonical five-scenario set at the nominal
operating point, $N{=}10$ seeds) of C2\_pid against C2 and against the C0
and C1 references, reported descriptively with bootstrap confidence
intervals and explicitly \textbf{without} an outcome-class classification.
The full off-nominal-grid evaluation of the PID-direct branch is future work
(deferred with the journal version): against a binding submission deadline
it would have doubled evaluation compute without bearing on the primary
contribution, whose outcome classification is defined on the MPC backend
only. Within its scope the demonstration still answers the practitioner's
threshold question --- whether agent-over-PID is \emph{operable} on this
plant class and where its performance sits relative to the MPC-backed
ladder on nominal-regime scenarios. Two boundaries of that answer are stated
in advance. First, the demonstration is nominal-regime only; it does not
test the off-nominal region where the headline results locate both C2's
strength (target acquisition) and its inverse finding. Second, the
disturbance\_rejection result from the primary ladder carries over as
deployment guidance in its own right --- steady-state disturbance rejection
at a known operating point is where a fixed-weight MPC outperformed the LLM
supervisor by an order of magnitude, i.e.\ \textbf{where LLM supervisors do
not belong} --- and removing the MPC does not remove that division of labor;
a practitioner reading this section should pair the agent with the
regulatory strategy each regime calls for, not expect the agent to
substitute for one.

The demonstration ran 2026-07-09 (50 cells, nominal OP $\times$ canonical
scenario set $\times$ 10 seeds). Data and execution logs are available
in the public project repository; see Table~\ref{tab:c2pid}.

\begin{table}[tbp]
 \centering
 \caption{Nominal-OP canonical IAE (mole-fraction$\cdot$min); C0/C1 baseline
 point references, C2/C2\_pid medians with bootstrap 95\,\% CIs. The
 \code{yD\_setpoint\_+0p5pct} row is C2-vs-C2\_pid comparable only (the
 baseline references track the setpoint step; the evaluation driver holds
 the operator spec --- the 4+1 scenario-set property,
 \S\ref{sec:limitations}).}
 \label{tab:c2pid}
 \small
 \begin{tabular}{lllll}
 \toprule
 Scenario & C0 (ref) & C1 (ref) & C2 med [95\,\% CI] & C2\_pid med [95\,\% CI] \\
 \midrule
 \code{F\_step\_+20pct} & 0.3312 & 0.0130 & 0.133 [0.117, 0.221] & 1.020 [0.924, 1.020] \\
 \code{F\_step\_-20pct} & 0.3038 & 0.0206 & 0.198 [0.128, 0.207] & 0.549 [0.501, 0.580] \\
 \code{zF\_step\_+10pct} & 0.0180 & 0.0118 & 0.0038 [0.0038, 0.0038] & 0.139 [0.105, 0.184] \\
 \code{zF\_step\_-10pct} & 0.0179 & 0.0056 & 0.0131 [0.0117, 0.0150] & 0.147 [0.110, 0.236] \\
 \code{yD\_setpoint\_+0p5pct} & n/c & n/c & 0.000 [---] & 0.000 [---] \\
 \bottomrule
 \end{tabular}
\end{table}

The threshold question is answered affirmatively: agent-over-PID is
\textbf{operable} on this plant class --- all 50 cells complete without a
failure, confirming the pre-registered operability expectation. On
performance, C2\_pid sits at the bottom of the ladder as pre-registered
(C2\_pid $\geq$ C2-over-MPC on every disturbance arm, disjoint CIs, factors
2.8--37$\times$) --- and \textbf{below C0 on every disturbance arm}, a
finding beyond the pre-registered expectation: at the nominal operating
point the ungated v1 agent over the PID stack is worse than the PID stack
alone with fixed setpoints. This is the \S\ref{sec:gatediscussion} nominal
mechanism (actively harmful v1 proposals, there corrected by the gate)
reappearing unclamped over a weaker regulatory layer; it sharpens, rather
than softens, the deployment guidance above.

\subsection{Transfer and deployment integration}\label{sec:transfer}

Transfer here is architectural, not empirical. Nothing in the two-layer
design is distillation-specific: the supervisory abstraction
(observe a regulated process at a slow cadence, propose setpoints, validate,
actuate through an unchanged regulatory layer) and the safety-gate mechanism
(fork a process twin, integrate the proposal forward, test against a pinned
constraint list, substitute on violation) require only three assets any
regulated process with a dynamic model possesses --- a twin, a constraint
envelope, and a setpoint interface. The chemistry here is the demonstration
vehicle; the methodology is \emph{designed to} transfer to coupled
multivariable processes in
semiconductor manufacturing (chamber thermal/pressure regulation), pharma
(bioreactor and crystallization control), battery production (formation and
electrode-coating lines), HVAC, and water treatment --- but \textbf{empirical validation on a second process class is future work,
not a claim of this paper}. Those sectors are listed as targets only,
\textbf{without claiming empirical coverage}
(\S\ref{sec:limitations}): the single-column scope is a disclosed
limitation, and the co-design findings of \S\ref{sec:gatediscussion}
(specification/envelope/admission-rule interaction) are exactly the class of
result a transfer study must re-establish per process.

On the deployment side, the evaluation stack uses open weights and an
OpenAI-compatible inference protocol so the same supervisor can be served
from a local or hosted endpoint without code change. The reproducibility
chain (\S\ref{sec:repro} --- pinned templates and sampling, no payment or
NDA gating of the method itself) is what makes that path auditable rather
than aspirational. Product-specific edge or digital-twin platforms are
integration choices, not claims of this paper.

\subsection{Future work: regime-aware hybrid supervision}\label{sec:hybridfw}

The regime map (\S\ref{sec:bucket}--\S\ref{sec:failuregeo}) points to a
next architecture that is \emph{not} evaluated end-to-end here: keep linear MPC as the default local regulator, and
invoke the LLM supervisor only when linearization drift or an off-nominal
target-acquisition task is detected, with the counterfactual gate remaining
mandatory on every LLM proposal. That hybrid is the deployment reading of ``where LLM supervisors do and do
not belong.'' Building and scoring it on the same 16-OP contract is left to
a successor study (together with optional
gain-scheduled or nonlinear-MPC baselines that would further pressure the
agent's target-acquisition advantage).

\subsection{Limitations and threats to validity}\label{sec:limitations}

\paragraph{Plant.}
LV RGA$(1,1)\approx 36$ (retained for literature comparison and fixed
regulatory layer). Fixed-gain C0 does not extrapolate to $F\pm 20\,\%$
(expected single-OP property). $\mathrm{cond}(G_{mv})$ grows $45\times$
toward $F{=}0.8$; target acquisition also collapses at $F{=}1.1$ from
nominal MVs. SIMC 2DoF is active on tracking arms but aggregate IAE does not
separate on a disturbance-dominated shootout mix \cite{skogestad2003simple}.

\paragraph{Model and baselines.}
Headline gate results are DeepSeek-V4-Flash-conditional
(Appendix~\ref{app:records}; \S\ref{sec:crossfamily}). Super reproduces the
DR \code{fails} band and failure geography with model-conditional magnitude;
TA strong cells are survivorship-qualified only ($33/800$ done). Family,
sampling temperature, and reasoning realization are confounded by per-model
DoE pins and provider templates ($T{=}0.3$ realized-off Super vs $T{=}1.0$
realized-on V4-Flash; \S\ref{sec:llmspec}). No gain-scheduled or nonlinear
MPC is in the ladder---C2's TA advantage is relative to the frozen
linear-MPC Pareto reference only.

\paragraph{Protocol.}
As executed, the canonical set is four feed disturbances plus one undisturbed
control arm (setpoint step held at operator spec). Single column; transfer
targets listed without empirical coverage; cross-domain detector not run;
real-plant validation deferred. C3-vs-C2 comparisons are distribution-level
(no seed pairing at $T{=}1.0$). Peak-deviation figures use the 5-min grid
and underestimate intra-cycle peaks ($\approx 2.7\times$ example).

\paragraph{Gate numerics.}
Violation uses $\varepsilon=10^{-6}$ on signed margins, adopted mid-evaluation
with documented, result-independent, direction-blind criteria; thresholds and
raw margins unchanged (Appendix~\ref{app:records}).

\section{Conclusion}\label{sec:conclusion}

Open-weight proposals only execute through a rule-based counterfactual gate
with named constraints and logged margins. That is the auditable surface.
The four-way ladder then produces an asymmetric headline: strong on
off-nominal target acquisition (IAE ratio $0.361$ at the upper CI against
Pareto-tuned linear MPC), fails on steady-state disturbance rejection
(upper-CI factor $16.03$; point $10.18$)---where LLM supervisors do not
belong. The gate compresses the worst failure mode into a bounded offset
($d\approx -1.4$) and makes the co-design problem concrete: $534$ of $590$
interventions are spec-on-bound geometry, even while $318$ blocks still stop
actively harmful nominal proposals and the best-behaved OP becomes
inoperable under a zero-margin specification.

None of the supervisory loop is distillation-specific. A twin, a constraint
envelope, and a setpoint interface are enough to port the pattern. That is
architecture, not multi-plant evidence. The nominal MPC-free branch is
operable but ranks below PID-only---another reason not to treat the agent as
a free replacement for industrial APC. Nemotron-3-Super keeps the
disturbance-rejection \code{fails} band and the plant-side failure geography;
magnitudes and protocol operability stay model-conditional, and Super
target-acquisition ``strong'' cells are survivors only
(\S\ref{sec:crossfamily}).

Practice: keep MPC for local regulation, use a gated LLM for re-planning,
and write the specification with the safety envelope in the same pass. A
closed-loop hybrid that implements that split is future work
(\S\ref{sec:hybridfw}); real-plant advisory validation is a separate track
(Appendix~\ref{app:records}).

\section*{Acknowledgments}
AI coding assistants were used under the author's direction for software
implementation; the author is solely responsible for the research design,
results, and manuscript.

\appendix

\section{Model-selection journey (audit trail)}\label{app:modelselection}

The initially selected primary was Nemotron-Super-49B-v1.5, chosen for its
agentic post-training lineage and workstation deployability. Two
reproducible failure modes surfaced on the local 8-bit MLX deployment: a
systematic over-purification bias and JSON truncation under
output-discipline pressure. Migrating the same model to hosted
full-precision inference (NVIDIA NIM) resolved the output-discipline failure
but not the underlying multi-step-coherence failure, isolating the defect
from the serving stack. A three-model comparison at the identical,
unmodified system prompt then located the failure as
capacity-tier-specific: on the nominal-baseline scenario the 49\,B model
produced canonical IAE 1.234, while Nemotron-3-Super-120B-A12B and
DeepSeek-V4-Flash both produced 0.000. The primary was upgraded to the
120\,B model and the ablation pinned to V4-Flash accordingly --- before any
off-nominal evaluation ran, and with the intermediate prompt-engineering
variants shelved once the multi-model evidence showed the issue was
capacity, not prompt. The journey is reported rather than compressed because it doubles as a negative result readers deploying local sub-100\,B
models will want: multi-step supervisory coherence was the binding
constraint, and it did not announce itself in single-shot smoke tests.

\section{DoE methodology and response surface}\label{app:doe}

Hyperparameter selection followed a screen-then-confirm pattern. Screening:
full-factorial temperature $\times$ top\_p $\times$ reasoning-configuration
($5 \times 3 \times 3 = 45$ cells), $N{=}5$ seeds per cell, 225 runs on the
nominal-baseline scenario, scored by canonical IAE against the
pre-registered 0.01 acceptance threshold. The response surface shows three
regimes: a plateau of near-zero IAE across $T \in \{0.0, 0.3, 0.6\}$ with
reasoning off; a cliff at $T{=}0.8$; and instability at $T{=}1.0$ in
non-reasoning mode that enabling chain-of-thought (budget 4096) rescues back
to the plateau --- a methodologically interesting observation in its own
right, since explicit reasoning here functions as a high-temperature
stabilizer and gives the architecture robustness headroom in two directions.
Confirmation: five additional seeds at the screening optimum ($N{=}10$
total), mean canonical IAE $5.75 \times 10^{-7}$ mole-fraction$\cdot$min
with all ten seeds identical to seven significant figures --- effectively
deterministic behavior at the pinned operating point on this scenario. The
selection of $T{=}0.3$ over the project-historical $T{=}0.6$ and over greedy
$T{=}0.0$ is a margin argument: $T{=}0.3$ sits in the plateau interior with
the largest distance to the $T{=}0.8$ cliff, while $T{=}0.0$ carries known
degenerative-repetition risk on long sequences.

\section{Hosted-endpoint precision (NVFP4 native)}\label{app:nvfp4}

Nemotron-3 Super was pre-trained in NVFP4 --- NVIDIA's 4-bit floating-point
format --- from the first gradient update; NVFP4 is the model's native
precision, and BF16 is the convenience up-cast, not the reference. The
hosted NIM endpoint serves the model at NVFP4. This inverts the usual
quantization caveat: our evaluation used the precision the model was trained
in, which is the \emph{most} faithful reproduction of its published
behavior, not a degraded approximation of it. Self-hosted replication should
use the published \code{-NVFP4} weight variant (natively on Blackwell-class
hardware, or via software emulation on Hopper-class hardware under vLLM);
replicators who instead load the BF16 variant should expect behavior
consistent with, but not bit-identical to, the hosted path.

\section{Decision Records and Reproducibility}\label{app:records}

\subsection*{A4.1 Decision records}
The design and scoping decisions cited in the main text by identifier
resolve to timestamped decision records in the project repository. Each was
recorded at the date shown; where the main text refers to a record as
\emph{frozen prior to} a later analysis, that temporal order is the record's
evidentiary content.

\begin{center}
\scriptsize
\setlength{\tabcolsep}{3pt}
\begin{tabular}{@{}l p{2.4cm} p{2.4cm} p{5.4cm}@{}}
\toprule
Record & Title & Date & Content \\
\midrule
ADR-003 & Novelty positioning & 2026-05-26 & Safety-gated agentic control as contribution \\
ADR-006 & Hierarchical control & 2026-05-26 & Supervisory agent over fixed regulatory layer \\
ADR-007 & Plant selection & 2026-05-26 & Skogestad Column~A (not a train) \\
ADR-008 & Deferred real-plant validation & 2026-05-27 & Anonymized industrial validation, separate track \\
ADR-009 & Regulatory backend & 2026-07-02 & MPC primary; PID as deployment-economics branch \\
ADR-010 & Fail-fast policy & 2026-05-28 & No silent fallback on inference/tooling errors \\
ADR-011 & Inference-stack split & 2026-06-01; Sub-Am.\ 2026-06-10 & Host split; Novita/V4 realized-on vs request-off \\
ADR-012 & Cross-family ablation & 2026-06-01 & DeepSeek as second-model ablation family \\
ADR-015 & Headline classification & 2026-07-02 & Outcome class on 16-OP grid; V4-Flash-scoped, frozen \\
ADR-016 & Safety-layer descope & 2026-07-02 & Counterfactual gate primary; detector timeboxed \\
ADR-017 & Cross-family re-instantiation & 2026-07-15 & Full Super (Nemotron-3) headline sweep \\
\bottomrule
\end{tabular}
\end{center}

\subsection*{A4.2 Repository}
All decision records, run configurations, per-cell artifacts, and results
memos referenced in this paper are versioned in the public project
repository
\url{https://github.com/cgncro-cyber/IndustrialAI}
under \code{docs/decisions/}, \code{docs/analyses/},
\code{data/reference/}, \code{data/case\_studies/}, and \code{data/runs/};
the descriptive references in the main text resolve to the identifiers
below. Large run trees may be partial in a shallow clone; the frozen
analysis memos under \code{docs/analyses/} are the citation-stable
summaries.

\subsection*{A4.3 Model serving identifiers}
The \textbf{headline} supervisor model, DeepSeek-V4-Flash, was served under
\code{deepseek/\allowbreak deepseek\allowbreak-v4\allowbreak-flash}
(OpenRouter$\rightarrow$Novita). The \textbf{cross-family} model,
Nemotron-3-Super-120B-A12B, was evaluated under
\code{nvidia/\allowbreak nemotron\allowbreak-3\allowbreak-super\allowbreak-120b\allowbreak-a12b}
(full Super headline on DeepInfra bf16; earlier screening also via NVIDIA
NIM).

\subsection*{A4.4 Run artifacts and pre-registration references}
\begin{center}
\footnotesize
\begin{tabular}{@{}p{4.4cm} p{8.2cm}@{}}
\toprule
Main-text reference & Repository path \\
\midrule
Pre-registered KPI specification & \code{docs/kpis.md} \\
Pre-registered outcome-class decision tree & \code{paper/\allowbreak methods\allowbreak\_phase3\allowbreak\_buckets.md} \\
C0 relay-tuning shootout data & \code{data/reference/\allowbreak c0\allowbreak\_pid\allowbreak\_tuning\allowbreak\_shootout.json} \\
Gate-evaluation case-study data & \code{data/\allowbreak case\allowbreak\_studies/\allowbreak c3\allowbreak\_gate\allowbreak\_eval\allowbreak\_2026-07-06/} \\
MPC-free demonstration run record & \code{docs/\allowbreak analyses/\allowbreak 2026-07-10\allowbreak\_c2\allowbreak\_pid\allowbreak\_demo\allowbreak\_results.md} \\
Cross-family Super headline results (frozen) & \code{docs/\allowbreak analyses/\allowbreak 2026-07-24\allowbreak\_xfam\allowbreak\_super\allowbreak\_headline\allowbreak\_results.md} \\
\bottomrule
\end{tabular}
\end{center}

\section{Supplementary: Nemotron-120B partial headline dataset}\label{app:supplementary}

\subsection*{S.1 Status and role of this dataset}

The Nemotron-120B 16-OP headline sweep (1600 cells; same grid,
scenarios, sub-metrics, seeds, and hardened analyzer pipeline as the
V4-Flash headline run) was executed on the NVIDIA NIM free-tier endpoint and
terminated twice at insufficient coverage: 419/1600 done on 2026-06-28, and
500/1600 done after a throughput-reduced gap-retry pass (2026-06-30
$\rightarrow$ 2026-07-02). Both terminations were dominated by provider-side
throttling, not by model or plant behavior. Per the pre-registered
patience-window rule, the cross-family-invariance claim was closed as
\textbf{unresolved-by-data} and the paper's headline classification is
\textbf{V4-Flash-only}; this supplementary documents the partial Nemotron
dataset so that (i) the covered operating points remain usable as
scope-limited cross-model evidence and (ii) the coverage decision is
auditable.

\subsection*{S.2 Coverage}

Final counts: \textbf{500 done / 1100 failed / 0 pending} (31.3\,\%
coverage).

\begin{table}[htbp]
 \centering
 \begin{tabular}{lll}
 \toprule
 $F$ column & Done / 400 & Note \\
 \midrule
 0.8 & 278 & LV-near-singular cluster covered; best-covered column \\
 0.9 & 64 & partial \\
 1.1 & 158 & partial; the hardest column of the V4-Flash grid \\
 \textbf{1.2} & \textbf{0} & entirely empty after two passes (800 consecutive failed attempts) \\
 \bottomrule
 \end{tabular}
\end{table}

Tuple-level coverage ((OP $\times$ scenario $\times$ sub-metric), $\geq 10$
done seeds $=$ the V4-Flash $N = 10$ protocol): \textbf{17/160} tuples
adequate; \textbf{79/160} at zero done. On the V4-Flash verdict-anchor OPs
($F{=}0.8$ LV-singular pair, $F{=}1.1$ column, $F{=}1.2$ column):
\textbf{10/100} tuples adequate; the three-corner-screening passive-follower-attractor
cluster $(F{=}1.2, z_F{=}0.55) \times \code{F\_step\_+20pct}$ is at 0/10 on
both sub-metrics.

{\sloppy
Failure classes of the 1100 failed cells (gap-retry pass):
\code{rate\_limit} 700 (63.6\,\%), \code{smoke\_parse} 147,
\code{server\_5xx} 140, \code{timeout} 92, \code{exit\_1} 16,
\code{exit\_2} 5. Host/transport/parse infrastructure
(\code{rate\_limit}+\code{server\_5xx}+\code{timeout}+\code{smoke\_parse}:
$700{+}140{+}92{+}147=1079$) is $1079/1100=98.1\,\%$ of failures; excluding
\code{smoke\_parse} ($700{+}140{+}92=932$) yields $932/1100=84.7\,\%$. An
earlier bare ``89.7\,\%'' figure is not used---it matches neither sum. The
$F{=}1.2$ emptiness is an order-vs-storm coincidence on both passes: the
driver reaches that column $\sim$33\,h into a pass, and the longest observed
healthy-endpoint stretch was 27.2\,h.\par}

\subsection*{S.3 What this dataset supports --- and what it does not}

\textbf{Supported (scope-limited to covered OPs):} cross-model observations
at the $F{=}0.8$ LV-near-singular cluster and the covered $F{=}1.1$ cells;
the cross-provider, cross-family reproducibility of the \emph{band
structure} asserted in the main text is additionally supported by the
Nemotron-120B three-corner screening (238/300 done; asymmetric
outcome-class-B signature: target\_acquisition strong, disturbance rejection
inverse), which remains the primary cross-model hint.

\textbf{Not supported:} any statement requiring the $F{=}1.2$ column, the
attractor cluster, or grid-level Nemotron aggregates --- in particular the
cross-family-invariance claim for the headline outcome-class-B classification.
Running the hardened analyzer on this dataset would produce a classification
structurally incomparable to V4-Flash at the verdict-anchor OPs; it was
therefore deliberately not produced.

\textbf{Scoping language (binding, from the pre-registration):} the
asymmetric outcome-class-B classification of V4-Flash is reported as \textbf{model-conditional}; the architectural-difficulty-zone observation
($F{=}0.8$ LV-singularity, $F{=}1.1$ hardest column) is a plant/baseline
property supported by V4-Flash data alone; the cross-model
architectural-limit comparison rests on the single overlapping OP
($F{=}0.8$/$z_F{=}0.55$: Nemotron 62\,\% vs V4-Flash 40\,\% failure) and
the three-corner-screening corners. A Nemotron-side verification of the prompt-disambiguation
sensitivity result was likewise not performed (no further NIM capacity for
this study) and is future work.

\bibliographystyle{unsrtnat}
\bibliography{references}

\end{document}

%% file: fig1_architecture_body.tex
%
\begin{tikzpicture}[
  font=\sffamily\small,
  configbox/.style={draw=black, thick, rounded corners=1.5pt, fill=white,
    minimum width=2.55cm, minimum height=1.05cm, align=center},
  layerlabel/.style={font=\sffamily\footnotesize\bfseries, anchor=north west},
  cadence/.style={font=\sffamily\footnotesize, anchor=north east, text=black!60},
  lane/.style={draw=black!45, fill=black!6, rounded corners=2pt},
  handoff/.style={-{Stealth[length=2.4mm]}, thick},
  note/.style={font=\sffamily\scriptsize, text=black!60, align=center},
]

\node[configbox] (c0) at (0,0) {\textbf{C0}\\ PID-only\\ (fixed setpoints)};
\node[configbox, right=0.45cm of c0] (c1) {\textbf{C1}\\ Linear MPC};
\node[configbox, right=0.45cm of c1] (c2) {\textbf{C2}\\ LLM agent};
\node[configbox, right=0.45cm of c2, minimum width=3.1cm] (c3)
  {\textbf{C3}\\ LLM agent\\ $\rightarrow$ safety gate};

\begin{scope}[on background layer]
  \node[lane, fit=(c0)(c3), inner xsep=8pt, inner ysep=14pt] (suplane) {};
\end{scope}
\node[layerlabel] at ($(suplane.north west)+(0.12,-0.10)$)
  {Supervisory layer \ (one configuration active per run)};
\node[cadence] at ($(suplane.north east)+(-0.12,-0.10)$) {cadence: 5--15 min};

\coordinate (mid) at ($(suplane.south)+(0,-1.15)$);
\node[draw=black, thick, minimum size=0.85cm, fill=white] (rl) at (mid) {};
\draw[thick] ($(rl.center)+(-0.28,-0.28)$) -- ($(rl.center)+(-0.06,-0.28)$)
  -- ($(rl.center)+(0.22,0.26)$);
\node[note, right=0.25cm of rl, anchor=west, align=left]
  {rate limiter\\ (setpoint ramp bound)};

\draw[handoff] (suplane.south) -- (rl.north)
  node[midway, left, note, align=right] {composition setpoints\\ $(y_D^{\mathrm{sp}},\,x_B^{\mathrm{sp}})$};

\node[configbox, minimum width=11.2cm, minimum height=1.0cm,
  below=2.35cm of suplane.south, anchor=north] (pid)
  {Regulatory multi-loop PID layer --- identical across C0--C3\\
   ($y_D$, $x_B$, condenser level $M_D$, reboiler level $M_B$; relay-feedback tuned)};

\begin{scope}[on background layer]
  \node[lane, fit=(pid), inner xsep=8pt, inner ysep=14pt] (reglane) {};
\end{scope}
\node[layerlabel] at ($(reglane.north west)+(0.12,-0.10)$) {Regulatory layer};
\node[cadence] at ($(reglane.north east)+(-0.12,-0.10)$) {cadence: 1--5 s};

\draw[handoff] (rl.south) -- (reglane.north);

\node[configbox, minimum width=11.2cm, minimum height=0.95cm,
  below=0.75cm of reglane.south, anchor=north, fill=black!12] (twin)
  {Process twin: Skogestad Column A (40-stage binary distillation, LV configuration)\\
   dynamic model integrated via \texttt{scipy.integrate}};

\draw[handoff] (reglane.south) -- (twin.north)
  node[midway, right, note] {MVs: reflux $L$, boilup $V$};
\draw[handoff] ($(twin.north)+(-3.2,0)$) -- ($(reglane.south)+(-3.2,0)$)
  node[midway, left, note] {measurements};

\draw[handoff, dashed]
  ($(reglane.west)+(-0.0,0)$) -- ++(-0.85,0) |- ($(suplane.west)+(0,0)$)
  node[pos=0.25, left, note, align=center, rotate=90, anchor=south]
  {observations (per supervisory cycle)};

\path[use as bounding box]
  ([shift={(-0.15cm,0.12cm)}]current bounding box.north west)
  rectangle
  ([shift={(0.15cm,-0.12cm)}]current bounding box.south east);

\end{tikzpicture}